\documentclass{aa}

\usepackage{graphicx}
\usepackage{natbib}
\usepackage[colorlinks,linkcolor=blue,anchorcolor=blue,citecolor=blue]{hyperref}
\hypersetup{
  colorlinks=true,
  allcolors=[rgb]{0,0,0.8},
  pdftitle={Rare Blue-core BCGs as Optical Signatures of ICM Cooling},
  pdfauthor={Kluge et al.},
  }
\usepackage{longtable}
\usepackage{pdflscape}
\usepackage{geometry}

\begin{document}
\title{Rare Blue-core BCGs as Optical Signatures of ICM Cooling}
%\subtitle{}

\author{M. Kluge\inst{1}\thanks{E-mail: mkluge@mpe.mpg.de},
E. Bulbul\inst{1,2},
X. Zhang\inst{1},
M. McDonald\inst{3},
% alphabetical
E. Artis\inst{1},
F. Balzer\inst{1},
R. Bender\inst{1,4},
Z. Ding\inst{1},
L. Fiorino\inst{1},
N. Malavasi\inst{1},
A. Merloni\inst{1},
T. Mistele\inst{1},
K. Nandra\inst{1},
M. Ramos-Ceja\inst{1},
M. Salvato\inst{1},
J. S. Sanders\inst{1},
S. Zelmer\inst{1}
}

\institute{
Max Planck Institute for Extraterrestrial Physics, Giessenbachstrasse 1, 85748 Garching, Germany \and
Faculty of Physics, Fakultät für Physik, Ludwig-Maximilians-Universität, Scheinerstr. 1, 81679 München, Germany \and
Kavli Institute for Astrophysics \& Space Research, Massachusetts Institute of Technology, Cambridge, MA 02139, USA \and
University-Observatory, Ludwig-Maximilians-University, Scheinerstrasse 1, D-81679 Munich, Germany
}

\date{Received XXX / Accepted XXX}
\titlerunning{Blue-core BCGs in eRASS:5}
\authorrunning{Kluge et al.}

\abstract{
Blue cores in brightest cluster galaxies (BCGs) provide an optical signature of recent central activity associated with cooling cluster atmospheres.
We search for these features in a $z<0.6$ subsample of clusters identified in five eROSITA all-sky surveys, combining eROSITA X-ray data, Legacy Surveys optical imaging, and optical spectroscopy.
We identify 379 blue-core BCGs among 11\,352 systems (3.3\%), although contamination increases toward higher redshift. Of these, 98 (0.86\% of the full sample) are X-ray bright.
Their defining signature is a central $g-r$ color at least $0.2$\,mag bluer than the reference population, typically extending to a radius $a\lesssim20\,{\rm kpc}$ before returning to the normal BCG envelope.
H$\alpha$ emission-line equivalent width correlates strongly with the blue continuum excess, consistent with young stellar components comprising $\sim$\,0.1--1\% of the pre-existing stellar mass.
However, emission-line-ratio diagnostics indicate widespread LINER-like excitation, with a stronger contribution from star formation in blue H$\alpha$ emitters but substantial non-stellar excitation.
This activity is closely linked to the thermodynamics of the host-cluster gas.
At $0.1R_{500}$, cooling times decrease from BCGs without significant H$\alpha$ emission to red and then blue H$\alpha$ emitters.
Blue BCGs preferentially occur in X-ray-luminous clusters at $L_{\rm X,300}\gtrsim10^{44}$\,erg\,s$^{-1}$, while red H$\alpha$ emitters are also biased toward higher X-ray luminosities.
The blue-core fraction increases with optical richness. 
For systems with $t_{\rm cool}(0.1R_{500})<20\,{\rm Gyr}$, doubling richness increases the odds of hosting a blue-core BCG by $2.45^{+0.39}_{-0.36}$. This dependence indicates that cluster scale carries information beyond cooling time, plausibly through the amount of gas available for cooling and condensation. Finally, the scarcity of blue BCGs without significant H$\alpha$ emission disfavors isolated short bursts followed by long H$\alpha$-quiet fading phases. It can instead be explained by sustained or closely recurrent star formation over timescales much longer than $100$\,Myr, or by H$\alpha$ persisting through additional excitation mechanisms as the young stellar component fades.
}


\maketitle

%----------------------------------------------------
\section{Introduction} \label{sec:intro}

The hot intracluster medium (ICM) dominates the baryonic content of galaxy clusters. With typical temperatures of $T\sim10^7$--$10^8\,{\rm K}$, the ICM emits primarily in X-rays. In relaxed clusters, the central ICM can form a dense cool core with a radiative cooling time below a gigayear. In the absence of heating, such short cooling times would imply large classical cooling flows, with mass-deposition rates of tens to thousands of solar masses per year \citep{Fabian1994}. However, high-resolution X-ray spectroscopy shows a strong deficit of emission from gas cooling to the lowest X-ray temperatures, and the observed amounts of cold gas and star formation in brightest cluster galaxies (BCGs) are usually much smaller than these classical predictions \citep{Peterson2006}. This discrepancy is the classical cooling-flow problem.

The current picture is that radiative cooling in cluster cores is regulated by feedback from the central active galactic nucleus (AGN). Mechanical energy from jets and buoyantly rising radio bubbles can offset a large fraction of the X-ray cooling losses, preventing most of the hot atmosphere from condensing into stars \citep{Churazov2001,Rafferty2006,McNamara2007,Fabian2012}. At the same time, feedback cannot be perfectly steady. Many rapidly cooling cluster cores contain multiphase gas, including optical line-emitting nebulae \citep{Crawford1999,Hatch2007,Hamer2016}, molecular gas reservoirs \citep{Edge2001,Salome2003,Russell2019}, and residual star formation in their central galaxies \citep{ODea2008,Donahue2010}.
These systems show that some fraction of the cooling ICM can still reach low temperatures and fuel activity in the BCG.

A useful way to describe this balance is precipitation-regulated feedback \citep{Voit2015}. In this framework, local thermal instability develops when the central ICM reaches sufficiently low entropy or short cooling time, allowing gas to condense out of the hot phase and form a multiphase medium \citep{Pizzolato2005, Sharma2012, Gaspari2012}. Observationally, star formation and optical line emission in BCGs are strongly enhanced in systems with low central entropy and short central cooling time, with commonly quoted thresholds of order $K_0\sim30\,{\rm keV\,cm^2}$ and $t_{\rm cool}\lesssim1\,{\rm Gyr}$ \citep{Cavagnolo2008, Rafferty2008, Voit2008, Donahue2010, Pulido2018}.
BCGs are therefore key laboratories for studying the connection between ICM cooling and galaxy growth. In most clusters, BCGs are massive, old, red galaxies that lie close to the center of the gravitational potential. However, a minority show optical emission lines \citep{Crawford1999,Edwards2007}, blue cores or ultraviolet excesses \citep{Bildfell2008,Pipino2009,Wang2010,Hicks2010}, or ongoing star formation \citep{Loubser2016}.
Such activity is preferentially found in BCGs closely aligned with the cooling ICM center as traced by the X-ray peak, with characteristic projected separations generally below $\sim$10--20\,kpc \citep{Bildfell2008,Rafferty2008,Sanderson2009}. This close alignment is characteristic of relaxed cool-core systems.

Short cooling times and close alignment between the BCG and the X-ray center nevertheless appear to be necessary but not sufficient conditions for star formation to occur \citep{Loubser2016}. BCGs embedded in apparently favorable cooling atmospheres can exhibit widely different levels of activity \citep{McDonald2019a}, ranging from nearly quiescent systems \citep{Somboonpanyakul2026} to extremely active objects such as the Phoenix cluster, whose BCG forms stars at a rate of $\sim$800\,$M_\odot\,{\rm yr^{-1}}$ while the surrounding ICM has a classical cooling rate of order $2000$--$4000\,M_\odot\,{\rm yr^{-1}}$ \citep{McDonald2012,McDonald2014}.

A fundamental open question is therefore what determines whether a favorable cooling atmosphere leaves the BCG largely passive, produces line-emitting multiphase gas, or proceeds far enough to actively form stars.
The outcome may depend not only on the current thermodynamic state, but also on the scale of the available gas reservoir and on when the system is observed within the cooling--feedback cycle. It also remains unclear whether the resulting star formation occurs in isolated short bursts or is sustained or recurrent over longer periods.

Previous results on the dependence of BCG activity on cluster scale are mixed. Globally, blue and star-forming BCGs have been found preferentially in lower-mass systems \citep{Pipino2011,Cerulo2019}, whereas strongly line-emitting BCGs become more common toward higher richness \citep{Liu2012}. Moreover, in strong cool cores, the BCG star-formation rate correlates with the absolute ICM cooling rate \citep{McDonald2018}, and the upper envelope of the classical mass-deposition rate increases with cluster mass \citep{Chen2007}. This raises the possibility that the overall scale of the cluster atmosphere contains information about BCG activity beyond the local cooling timescale alone. 

Large-area X-ray surveys now make it possible to search for rare phases of BCGs in statistically useful cluster samples. The extended ROentgen Survey with an Imaging Telescope Array (eROSITA; \citealt{Predehl2021}) on board the Spectrum-Roentgen-Gamma observatory (SRG; \citealt{Sunyaev2021}) has delivered large samples of X-ray-selected sources in the first and cumulative third eROSITA All-Sky Surveys (eRASS1; \citealt{Merloni2024} and eRASS:3; \citealt{RamosCeja2026}). The corresponding eRASS1 cluster catalog contains 12\,247 spatially extended X-ray sources identified as galaxy groups and clusters \citep{Bulbul2024}, with optical identification based on red-sequence galaxy overdensities \citep{Kluge2024}. In this work, we use the deeper eRASS:5 data set, corresponding to an effective depth of 4.4 all-sky surveys, to search for rare BCGs with anomalous central optical colors in X-ray-selected environments.

The paper addresses five main questions. First, we measure the observed incidence of centrally blue BCGs in an X-ray-selected parent sample. Second, we investigate the relation between the blue continuum excess and optical line emission and assess their consistency with recent star formation. Third, we examine whether the blue-core population remains preferentially associated with shorter-cooling ICM atmospheres despite a limited spatial resolution in the cluster cores. Fourth, we test whether cluster scale, traced by optical richness, predicts blue-core activity after accounting for the measured cooling time and redshift. Fifth, we use the relative occurrence of blue BCGs with and without H$\alpha$ emission to constrain the relation between the blue stellar and line-emitting phases and their characteristic visibility timescales.

The paper is organized as follows. 
Section \ref{sec:data} describes the data: the eRASS:5 cluster catalog, X-ray thermodynamic profiles, BCG identification, and the construction of the visual, statistical, and spectroscopic samples. Section \ref{sec:methods} describes our methods: the color-profile and emission-line measurements, starburst reference models, and the X-ray and statistical methods used in the analysis. Section \ref{sec:results} presents our results: the blue-core selection and contamination tests, spectroscopic diagnostics, X-ray properties of the host clusters, and the dependence of the blue-core incidence on redshift and cluster richness. Section \ref{sec:discussion} discusses the roles of ICM thermodynamics and cluster scale in producing blue stellar components, and the relation between the blue stellar and line-emitting phases in the context of feedback-regulated cooling. Section \ref{sec:conclusions} summarizes our conclusions. Throughout this paper we adopt a flat $\Lambda$CDM cosmology with $H_0=70\,{\rm km\,s^{-1}\,Mpc^{-1}}$, $\Omega_{\rm M}=0.3$, and $\Omega_\Lambda=0.7$, and quote magnitudes in the AB system.

%----------------------------------------------------
\section{Data and sample selection}\label{sec:data}

This section describes the parent cluster catalogs, measurement of thermodynamic ICM properties, two independent BCG selection methods, final BCG samples, and the spectroscopic data used in this work.

\subsection{Cluster catalogs}

Our parent sample is based on the eRASS:5 galaxy cluster catalogs that cover the western galactic hemisphere at Galactic longitude $179.9442\degr < l < 359.9442\degr$. X-ray source detections on 4.4 all-sky scans were performed using the standard eROSITA Science Analysis Software System \citep[eSASS;][]{Brunner2022} version 030.

Optical identifications of these cluster candidates are performed with the eROMaPPer pipeline \citep{IderChitham2020,Kluge2024} that is based on the red-sequence-based cluster finder redMaPPer \citep{Rykoff2012,Rykoff2014,Rykoff2016}.
The optical data used for cluster identification and our BCG photometry are provided by the DESI Legacy Imaging Surveys \citep{Dey2019}, which provide wide-area optical imaging over most of the extragalactic sky. We use the data release (DR) 10.1 products in the south and the corresponding northern imaging from DR9, following the eRASS1 optical-identification analysis of \citet{Kluge2024}. The common usable footprint of eRASS:5 and the LS is 13\,335\,deg$^2$.

Galaxy clusters are usually detected as extended sources in eROSITA data, but a non-negligible fraction is classified as point-like \citep{Balzer2025}. This is especially true for strong cool-core clusters, which exhibit bright, peaked X-ray surface brightness (SB) profiles in their cores, such as the Phoenix cluster \citep{McDonald2012}. We therefore include both source classes in our parent sample.

Following the quality criteria described in \citet{Kluge2024}, we require the sources to lie within the usable optical survey footprint ({\tt IN\_FOOTPRINT}) and below the local limiting redshift set by the imaging depth ({\tt IN\_ZVLIM}), exclude regions containing known non-cluster extended X-ray sources ({\tt IN\_XGOOD}), and remove split X-ray detections ({\tt SPLIT\_CLEANED}).
The cluster redshifts {\tt BEST\_Z} are constrained to $z<0.6$ to keep the H$\alpha$ emission line within the observed wavelength range of a typical spectrum from the Sloan Digital Sky Survey (SDSS). We further require a richness of $\lambda_{\rm norm}>20$, below which catalog contamination increases considerably \citep{Kluge2024}, and a cluster masking fraction of ${\tt MASKFRAC}<0.3$.
These catalog-level criteria together yield parent samples of 12\,889 clusters in the extended-source catalog and 14\,433 clusters in the point-source catalog.

Cooling-driven BCG activity is observed preferentially in systems with small BCG--X-ray offsets, motivating us to require the BCG to lie within 30\arcsec~of the X-ray detection. This conservative criterion is consistent with the half-energy width of the eROSITA point-spread function \citep[PSF;][]{Merloni2024}---although the astrometric precision is much higher---and is deliberately chosen to be broad compared with the $\sim$10--20\,kpc offsets typically observed in active cool-core systems (see Sec.~\ref{sec:intro}). 
The centering criterion keeps systems with projected BCG--X-ray offsets of $\lesssim100$\,kpc at $z\gtrsim0.2$. At a relative velocity of $1000$\,km\,s$^{-1}$, traversing this distance requires $\sim$100\,Myr, comparable to the minimum visibility time of the blue continuum, derived in Sec. \ref{sec:ha_tracks}. A merger that abruptly interrupted the cooling supply should therefore leave most post-starburst BCGs within our selection.

\subsection{X-ray data and thermodynamic profiles} \label{sec:xray_data}

We use eRASS:5 X-ray data from processing version 030. Images are generated in the 0.3--1.0, 1.0--2.3, and 2.3--7.0\,keV bands, together with vignetting-corrected exposure maps. The three energy bands are fitted simultaneously with \texttt{MBProj2D} \citep{Sanders2018}, following the general procedure described by \citet{Bulbul2024}. The model images are convolved with the eROSITA PSF before being compared with the observed images. The gas density is modeled using the first term of Eq.~3 of \citet{Vikhlinin2006}, with $\gamma=3$, while the ICM is assumed to be isothermal.
The metallicity is fixed to 0.3\,Z$_\odot$ using the abundance table of \citet{Asplund2009}, and Galactic absorption is modeled using \texttt{NHTOT} \citep{Willingale2013}. A spatially flat background component is fitted independently in each energy band.

Nearby X-ray sources are either masked or simultaneously modeled, following the procedure adopted for the eRASS:5 cluster analysis. Importantly, for clusters drawn from the point-source catalog, the central point-source detection is not masked because it corresponds to the cluster itself. Parameter posteriors are sampled with \texttt{emcee}; the chains are run for 5000 steps, and the first 1000 steps are discarded. 
From the sampled thermodynamic profiles, we calculate the entropy proxy as

\begin{equation}
K = kT n_{\rm e}^{-2/3},
\end{equation}
and the isobaric radiative cooling time as
\begin{equation}
t_{\rm cool}
=
\frac{5}{2}
\frac{(n_{\rm e}+n_{\rm i})kT}
     {n_{\rm e}n_{\rm H}\Lambda(T,Z)},
\end{equation}
where $n_{\rm e}$, $n_{\rm i}$, and $n_{\rm H}$ are the electron, ion, and hydrogen number densities, respectively, $\Lambda(T,Z)$ is the radiative cooling function at temperature $T$ and metallicity $Z$, and $k$ is Boltzmann's constant.
We use the resulting median gas-density, entropy, and cooling-time profiles in the following analysis.
Throughout this work, X-ray luminosities are measured in the 0.2--2.3\,keV band within a projected radius of 300\,kpc.

\subsection{Primary BCG selection} \label{sec:bcg1}

BCG identification can be ambiguous because luminosity, morphology, and proximity to the cluster center do not always identify the same galaxy (see App.~\ref{app:bcg} and e.g., \citealt{Kluge2020}).
To improve the identification in the eRASS:5 cluster catalog, we have trained a gradient-boosted decision tree ranker using the {\tt LambdaRank} objective, as implemented in {\tt LightGBM} \citep{Ke2017}. For each cluster, the ranker assigns a BCG score to the 50 brightest galaxies in the $z$ band within the richness-based empirical cluster radius $R_{\lambda}$ from the X-ray center \citep{Rykoff2014}. We explicitly include galaxies that are not classified as red-sequence members by eROMaPPer, because active BCGs with anomalous colors would otherwise be excluded.
The ranker combines information on galaxy position, luminosity, morphology, color, cluster membership, and local galaxy density. A SHapley Additive exPlanations (SHAP) analysis shows that the separation from the X-ray detection, $z$-band magnitude, and angular galaxy extent provide the strongest individual constraints, while galaxy colors carry less weight. The training data, full feature set, and corresponding SHAP diagnostics are presented in App.~\ref{app:bcg}.

This machine-learning-based approach recovers the manually chosen BCG in 93\% of cases on a training set that includes 80\% of the clusters. On a validation set of the remaining 20\% of clusters, the model achieves 88\% correct predictions. Most of the discrepant clusters contain a similar second BCG, making the choice intrinsically ambiguous. For the final assignment, we train the ranker on all manually selected BCGs.

\subsection{Alternative BCG selection} \label{sec:bcg2}

In extreme cases where the BCG color is far from the red sequence, such as the Phoenix cluster, the machine-learning-based BCG selection fails. Therefore, we add a secondary BCG identification where the BCG candidate is required to be brighter in the $z$ band than the brightest red-sequence member galaxy.
We adopt a more restrictive search radius of 15\arcsec~around the initial X-ray detection to reduce false-positive associations and preferentially select the most securely centered systems, motivated also by the close BCG--X-ray alignment observed in strong cool-core clusters \citep{Bildfell2008,Rafferty2008,Sanderson2009}.

This secondary procedure identifies 541 BCG candidates in the extended-source parent sample and 1968 in the point-source parent sample.
After removing candidates that are identical to those already identified by the primary selection, the secondary selection adds 190 (912) BCG candidates to the 9864 (5765) primary BCGs in the X-ray extended (point-like) source catalog, including the BCG of the Phoenix cluster.
Because 94 (189) clusters contain two distinct BCG candidates, the final well-centered parent cluster sample contains 9960 (6488) unique clusters.

\subsection{Visually selected active-BCG sample} \label{sec:init_visual}

The visually selected sample is used to identify the characteristic photometric signatures of active BCGs and to motivate the quantitative color-profile criteria that will be introduced in Sec.~\ref{sec:colorprofiles}.
We begin with all BCGs returned by the alternative BCG selection. That provides a broad sample of 2509 well-centered, strongly color-offset BCGs without preselecting the particular photometric signature that we aim to identify.
Each candidate was visually examined using $g,r,z$ color images from the Legacy Surveys, eRASS:5 X-ray contours, encircled member galaxies identified by eROMaPPer, and a cross indicating the active BCG candidate. The inspection was performed with an internal version of the eRASS Cluster Inspector.\footnote{\url{https://erosita.sciserver.mpe.mpg.de/cluster-inspector/erass1}}

This visual-inspection sample contains many background AGN in the point-source catalog and many foreground galaxies in the extended-source catalog. Without redshift information, we identify good candidates for active BCGs by surrounding intracluster light (ICL) colors that are visually consistent with the colors of the member galaxies, that is, likely belonging to the cluster, and having regions, usually the core, that deviate noticeably from that color. We obtain 50 good candidates in the extended-source catalog and 19 good candidates in the point-source catalog, including the well-known active BCGs in the Abell 1835 (extended source, primary BCG selection) and Phoenix clusters (point source, alternative BCG selection).

\subsection{Statistical BCG sample}
\label{sec:statistical_sample}

For the statistical analysis, we combine the four BCG-selection branches described above (primary and alternative BCG selections in the extended- and point-source catalogs) and, where a spectroscopic BCG redshift is available, require consistency with the photometric cluster redshift, $|z_{\rm spec,BCG} - z_{\lambda}|<0.06$. This leaves 16\,097 BCG candidates before resolving multiple assignments to the same cluster.

We additionally require valid SB and color profiles suitable for the quantitative classification described in Sec.~\ref{sec:colorprofiles}; hence, we refer to it also as the color-selected sample. In brief, this requires plausible SB measurements of the BCG and surrounding ICL. Individual clusters can occur in both the primary and alternative BCG-selection branches. We therefore retain only one BCG assignment per cluster. For blue-core candidates, the adopted BCG is determined by visual inspection; for the complementary control population, we preferentially retain the BCG identified by the primary machine-learning selection. After applying the stellar contamination rejection described in Sec.~\ref{sec:gaia}, the final statistical sample contains 11\,352 unique BCGs.

\subsection{Spectroscopic data and samples}
\label{sec:spectroscopic_data}

We assembled archival optical spectra from DR19 of the Sloan Digital Sky Survey (SDSS; \citealt{SDSSCollaboration2026}) and DR1 of the Dark Energy Spectroscopic Instrument (DESI; \citealt{DESICollaboration2026}), supplemented by dedicated observations with the Hobby--Eberly Telescope \citep[HET;][]{Ramsey1998,Hill2021} Low Resolution Spectrograph~2 \citep[LRS2;][]{Chonis2014,Chonis2016}.

For SDSS, we queried the DR19 Valis application programming interface at the position of each BCG and associated the nearest spectroscopic target within $3\arcsec$. We retrieved the corresponding coadded spectrum, preferentially using the all-epoch coadd, together with the wavelength-dependent instrumental resolution required for the subsequent spectral analysis.

For DESI, we queried the \texttt{desi\_dr1.zpix} catalog through the NOIRLab Astro Data Lab Table Access Protocol service and selected the nearest target within $3\arcsec$ of the BCG using the mean fiber coordinates. We retrieved the corresponding public DR1 \texttt{iron} HEALPix coadd and combined the B, R, and Z cameras into a single wavelength-ordered spectrum. DESI covers approximately $3600$--$9800$\,\AA\ at a wavelength-dependent resolving power of $R\simeq2000$--5000 \citep{DESICollaboration2022}.

To supplement the archival spectroscopy, we obtained grey-time HET/LRS2 observations of 23 objects selected from the visual sample (Sec. \ref{sec:init_visual}) based on observability.
LRS2 is a fiber-fed integral-field spectrograph whose blue- and red-optimized units together cover approximately $3700$--$10500$\,\AA\ at resolving powers of $R\sim1100$--1900. Total exposure times ranged from 1800 to 3600\,s. The data were reduced with the Panacea pipeline\footnote{\url{https://github.com/grzeimann/Panacea}}, including bias subtraction, cosmic-ray masking, fiber extraction, wavelength and flux calibration, and correction for atmospheric differential refraction. We use the calibrated data cubes of the target and sky pointings.

To exclude objects with implausible surrounding ICL, we require objects to fall into either the blue or red photometric classification defined in Sec. \ref{sec:colorprofiles}. The Gaia foreground-star rejection described in Sec. \ref{sec:gaia} is applied to the blue subsamples. The resulting full spectroscopic sample contains 2349 unique BCGs with usable spectral measurements. In cases of coverage by multiple instruments, we adopt the spectrum with the highest continuum signal-to-noise ratio (SNR), resulting in 858 DESI, 1489 SDSS, and two HET/LRS2 spectra. For comparisons based on the presence or absence of significant H$\alpha$ emission, we additionally apply the spectroscopic classification criteria defined in Sec. \ref{sec:fitspec}. The resulting H$\alpha$-classified spectroscopic sample contains 1130 BCGs: 75 with significant H$\alpha$ emission and 1055 without significant H$\alpha$ emission, comprising 383 DESI, 746 SDSS, and one HET/LRS2 spectrum.

\section{Methods} \label{sec:methods}

\subsection{Color profiles}\label{sec:meth:colorprofiles}

We queried $600\times600$ pixel cutouts with the native resolution of 0.262\arcsec\,pixel$^{-1}$ from the Legacy Surveys in the $g$, $r$, $i$, and $ z$ filter bands, centered on the BCG.\footnote{For the images, we query the url \url{https://www.legacysurvey.org/viewer/cutout.fits?ra={RA_BCG}&dec={DEC_BCG}&layer=ls-dr10&pixscale=0.262&size=600}}
To measure unbiased color profiles, we must ensure consistent sampling of the galaxy structure. Therefore, we use identical masks, point-spread functions (PSFs), and isophotes across the different filter bands. Each image was convolved with an individual kernel to match its PSF to that of the image with the worst seeing \citep{Goessl2002}. The PSFs are also queried from the Legacy Surveys.\footnote{For the PSF models, we query the url \url{https://www.legacysurvey.org/viewer/coadd-psf/?ra={RA_BCG}&dec={DEC_BCG}&layer=ls-dr10}}
We caution that the PSF cutout size in the northern Legacy Survey's $g$ and $r$-band data is undersampled by a factor of two. We upsample them using a 3rd-order 2-D spline interpolation.

Next, we generate source masks for each filter band individually, combine them into a single mask, and apply it to the images across all available filter bands.
To measure the BCG SB profiles, we apply the methodology described in \citet{KlugeBender2023}. The procedure masks non-BCG sources using multiple SB thresholds after 2-D spline removal of extended structures such as ICL, PSF wings, or other background inhomogeneities \citep{Kluge2020,Kluge2020T}. Afterward, elliptical isophotes are determined using the {\tt python} package {\tt photutils} \citep{Bradley2023} on the $z$-band images.
Once the elliptical shape parameters become unstable, we fix them and only increase the semi-major axis radius $a$.
In the inner 15 pixels, we use the surface fluxes measured by {\tt photutils} through interpolated sampling along fitted elliptical isophotes. Beyond this radius, we use the median unmasked surface flux within annuli around the isophotes to optimize the SNR.
The background in the Legacy Surveys images is already close to zero. In our tests, not subtracting an additional constant background proved more stable than estimating one individually around each BCG.
The background in the northern $z$ band had been oversubtracted, causing the SB and color profiles to decline unphysically steeply. This is easily seen in the Legacy Viewer by the distinctly cyan-colored halos around nearby extended galaxies. Hence, we keep the $z$ band only in the southern LS. As we will see later, the distinguishing color for active BCGs is $g-r$, which is unaffected by this oversubtraction.

The SB profiles were corrected for Galactic extinction using the Schlegel, Finkbeiner \& Davis dust map \citep{Schlegel1998}, with the Legacy Surveys extinction coefficients $A_\lambda/E(B-V)=3.214,2.165,1.592,1.211$ for the $g,r,i,z$ bands, respectively \citep{Schlafly2011}. We additionally K-corrected each band to the corresponding rest-frame band using \texttt{EzGal} \citep{Mancone2012}. The K-corrections were calculated using the stellar-population synthesis models of \citet{Bruzual2003}, assuming an exponentially declining star-formation history, solar metallicity, the initial mass function of \citet{Salpeter1955}, and a formation redshift of $z_{\rm f}=3$. We used the DECam $g,r,i,z$ filter transmission curves. The same K-correction model and Galactic extinction value were applied across all radii of a given BCG. SB profiles were further corrected for cosmological SB dimming by subtracting $10\log_{10}(1+z)$.

Finally, the corrected SB profiles were subtracted from one another to obtain rest-frame color profiles. Uncertainties are not estimated because outliers will be identified using the scatter of the measurements across thousands of color profiles.

\subsection{Spectral fitting and emission-line measurements}\label{sec:fitspec}

All SDSS, DESI, and HET/LRS2 spectra were uniformly re-analyzed using the workflow described in \citet{KlugeBender2023}. While the original implementation focused on stellar kinematics and absorption-line indices, here we use it to obtain homogeneous continuum-subtracted emission-line measurements.
For the stellar continuum, we use the X-shooter stellar library DR2 \citep{Arentsen2019,Gonneau2020}. The UVB templates are used for the blue part of the spectrum ($3000$\,\AA$\ <\lambda_{\rm rest}<5500$\,\AA) and the VIS templates for the red part ($5500$\,\AA$\ <\lambda_{\rm rest}<10\,000$\,\AA) in two separate fits.

The stellar continuum and emission lines are fitted simultaneously with the Penalized Pixel-Fitting method \citep[pPXF;][]{Cappellari2004,Cappellari2017} after matching the stellar templates to the instrumental resolution. The continuum is fitted without additive or multiplicative polynomials to avoid biasing the emission-line fluxes. Narrow emission lines are modeled by Gaussian templates, while an additional broad Gaussian Balmer component is included where required. The theoretical 3:1 amplitude ratios are tied for the [O~{\sc iii}], [O~{\sc i}], and [N~{\sc ii}] doublets. Line fluxes and rest-frame equivalent widths (EWs) are measured from the fitted gas-template amplitudes. Full details of the fitting procedure and uncertainty calculation are given in App.~\ref{app:spectral_fitting}.
Based on visual inspection of the spectral fits, we define a significant H$\alpha$ detection by requiring a fitted EW(H$\alpha$) SNR greater than 7 and a gas velocity dispersion of $\sigma_{\rm H\alpha}>60\,{\rm km\,s^{-1}}$ to mitigate false detections of sky-line subtraction residuals.
Conversely, among spectra covering the redshifted H$\alpha$ wavelength, we classify those with a continuum SNR above 10 and an H$\alpha$ SNR below 4 as having no significant H$\alpha$ detection.

\subsection{Starburst reference models} \label{sec:starburst_models}

To provide a simple reference for the joint evolution of the optical continuum and H$\alpha$ emission, we compute instantaneous-burst tracks using the Flexible Stellar Population Synthesis code (FSPS; \citealt{Conroy2009,Conroy2010}) accessed through Prospector \citep{Johnson2021}. We use simple stellar populations based on MIST isochrones \citep{Choi2016,Dotter2016} and the MILES stellar library \citep{SanchezBlazquez2006,FalconBarroso2011}, assuming a Salpeter initial mass function \citep{Salpeter1955}. Nebular line and continuum emission are computed following \citet{Byler2017}, as implemented in FSPS using Cloudy models.

We adopt a 10\,Gyr single stellar population as the passive BCG component and add a young single stellar population of varying age. Separate tracks are computed for burst mass fractions $f_{\rm burst}=0.03\%,\,0.1\%,\,0.3\%$, and $1.0\%$. The old and young components are mixed by stellar mass. Synthetic rest-frame DECam-like $g,r,i,z$ broadband fluxes are computed at $z=0$, including the stellar and nebular continua but excluding nebular emission lines. This provides a single redshift-independent reference for the continuum evolution, whereas the broadband contribution of strong emission lines depends on redshift as they move between the observed-frame filters. In particular, H$\alpha$ shifts out of the DES $r$ band at $z\simeq0.08$ and therefore does not affect the observed $g-r$ color, our clearest color diagnostic, for the large majority of the sample.

From these models we derive the continuum colors, $r$-band brightening, and H$\alpha$ equivalent width EW(H$\alpha$). The latter is calculated from the FSPS H$\alpha$ line luminosity divided by the mixed continuum flux density at 6563\,\AA. The $\mu_r$ tracks are normalized to the median observed 8\,kpc SB of H$\alpha$-weak BCGs. Likewise, each color track is anchored to the corresponding median color of the H$\alpha$-weak BCG population, while retaining the FSPS-predicted color change relative to the 10\,Gyr passive population.

\subsection{Radial X-ray measurements of cooling time and entropy} \label{sec:xray_methods}

We use the median entropy and cooling-time profiles derived from the eRASS:5 \texttt{MBProj2D} analysis described in Sec.~\ref{sec:xray_data}. Although the eROSITA PSF is explicitly included in the forward modeling, the finite spatial resolution limits the information available on scales below the PSF. Values at the innermost model-grid point are therefore not necessarily directly constrained at the corresponding physical scale. This is particularly important for entropy and cooling time, which generally decrease toward the cluster center \citep{Bauer2005,Donahue2006,Hogan2017,Main2017}: when the inner profile is weakly constrained, the innermost value can therefore be sensitive to the adopted profile model and inward extrapolation.

We consequently evaluate the thermodynamic profiles at radii where these systematic effects are reduced, while remaining as close to the core as resolution permits. We assess this at the population level by examining the dependence of the median cooling time on redshift and richness.
Small radii are increasingly sensitive to the finite spatial resolution and inward profile extrapolation, whereas fixed physical radii introduce a stronger richness-dependent aperture effect. We assess radial systematics by comparing cooling-time measurements at 50, 100, and 150\,kpc and at $0.05$, $0.10$, and $0.15R_{500}$.

For the scaled radii, $R_{500}$ is estimated from the optical richness $\lambda_{\rm norm}$ and cluster redshift by inverting the eRASS1 richness--mass relation of \cite{Ghirardini2024}
\begin{equation}
\begin{split}
\ln \lambda_{\rm norm} ={}& \ln A_\lambda
+ \left[B_\lambda
+ D_\lambda \ln\!\left(\frac{1+z}{1+z_{\rm piv}}\right)\right] \times \ln\!\left(\frac{M_{500}}{M_{\rm piv}}\right) \\
&+ C_\lambda \ln\!\left(\frac{1+z}{1+z_{\rm piv}}\right),
\end{split}
\end{equation}
with $\ln A_\lambda=3.08$, $B_\lambda=1.01$, $C_\lambda=-0.49$, $D_\lambda=0.30$, $M_{\rm piv}=2\times10^{14}\,M_\odot$, and $z_{\rm piv}=0.35$. We then compute

\begin{equation}
  R_{500}=\left[\frac{3M_{500}}
  {4\pi\,500\rho_{\rm crit}(z)}\right]^{1/3}.
\end{equation}

The full comparison is presented in App.~\ref{app:xray_radius}.
We adopt $0.1R_{500}$ as our fiducial radius and 100\,kpc for a complementary robustness test.

\subsection{Statistical model for the richness dependence of blue-core incidence} \label{sec:statistical_models}

Blue-core incidence depends strongly on the thermodynamic state of the cluster core, and may also vary with cluster richness and redshift. We therefore ask whether, at fixed measured cooling time and redshift, richer clusters are more likely to host a blue-core BCG. To quantify this, we fit a multivariable logistic regression to the color-selected blue and red BCG populations.
Because the BCG classification is binary, we use logistic regression, with the logit transforming the blue-core probability from $0<P<1$ to an unbounded quantity that can be modeled as a linear function of richness, cooling time, and redshift:

\begin{equation}
\begin{aligned}
\operatorname{logit} P({\rm blue})
&\equiv
\ln\!\left[\frac{P({\rm blue})}{1-P({\rm blue})}\right] \\
&=
a + b_\lambda \log_{10}\!\left(\frac{\lambda_{\rm norm}}{50}\right)
+ b_t \log_{10}\!\left(\frac{t_{\rm cool}}{10\,{\rm Gyr}}\right) \\
&\quad
+ b_z \frac{z-0.3}{0.1}.
\end{aligned}
\label{eq:logistic_blue}
\end{equation}
Here, $b_\lambda$, $b_t$, and $b_z$ describe the dependence of the log-odds on richness, cooling time, and redshift, respectively. The predictors are centered at $\lambda_{\rm norm}=50$, $t_{\rm cool}=10$\,Gyr, and $z=0.3$, representative values for the fitted sample.
In this parameterization, $b_\lambda>0$ implies that, at fixed $t_{\rm cool}$ and redshift, the odds of hosting a blue-core BCG increase with cluster richness. The corresponding odds ratio for a factor-of-two increase in richness is
\begin{equation}
{\rm OR}_{\lambda\times2}
= \exp\!\left[b_\lambda \log_{10}(2)\right].
\end{equation}

Our primary comparison is restricted to systems with $t_{\rm cool}<20$\,Gyr at the tested radius. This cut empirically isolates the shorter-cooling part of the population and is not intended as a conventional cool-core threshold. The subsample was chosen to focus the primary inference on the cooling-time regime where both blue-core and red BCGs are represented. We perform two complementary robustness tests. First, we repeat the linear model using the full sample with valid cooling-time measurements, without imposing the 20\,Gyr limit, to test whether the inferred richness dependence is induced by this selection. Second, within the shorter-cooling sample, we add a quadratic term $b_{t,2}[\log_{10}(t_{\rm cool}/10\,{\rm Gyr})]^2$ to Eq.~\ref{eq:logistic_blue} to test sensitivity to the assumed functional form of the strong cooling-time dependence. 
There may be a rapid transition toward short cooling times, saturation, or threshold-like behavior. If the model were forced to follow $\operatorname{logit} P(\rm blue) \propto \log(t_{\rm cool})$ when the true dependence were curved, residual $t_{\rm cool}$ structure could correlate with richness and masquerade as a richness effect.
Uncertainties in the regression coefficients and derived odds ratios are estimated from 2000 bootstrap resamples.

\section{Results} \label{sec:results}

\subsection{Visual properties of active-BCG candidates}

\begin{figure*}
    \centering
    \includegraphics[width=\linewidth]{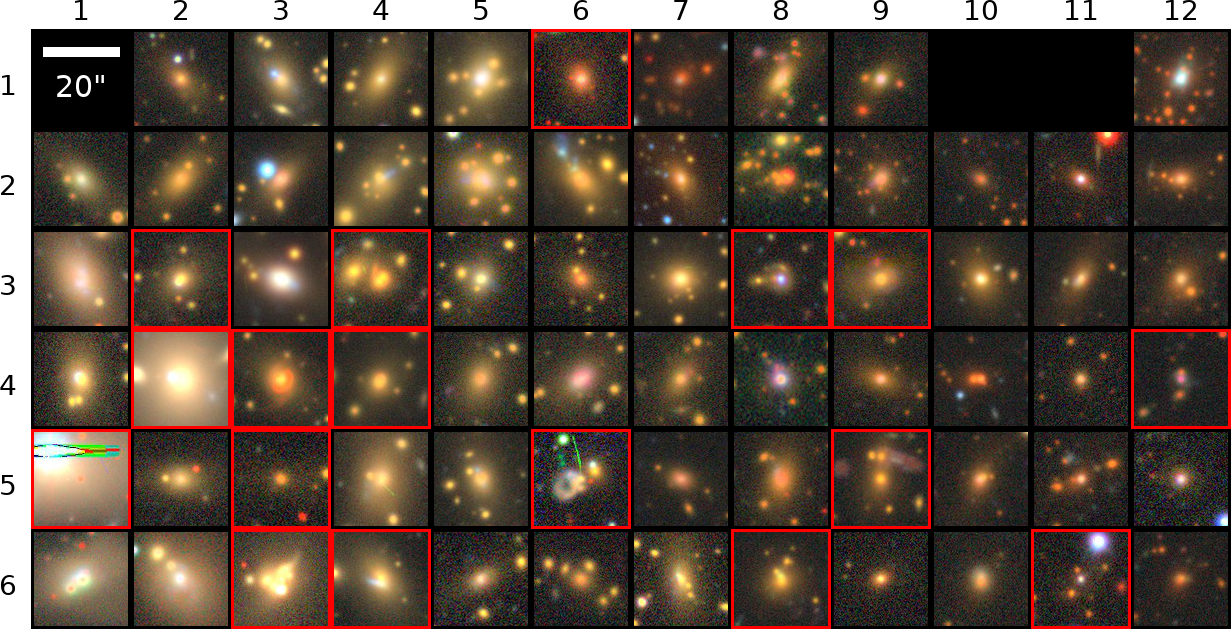}
    \caption{Visual sample of active-BCG candidates. Each panel shows a Legacy Surveys observed-frame $grz$ cutout centered on the candidate BCG. Candidates are ordered by increasing redshift from left to right and by increasing richness from bottom to top. Red frames mark objects rejected by the quantitative SB and color-profile cuts.}
    \label{fig:visualsample}
\end{figure*}

Figure~\ref{fig:visualsample} shows the observed-frame $grz$ images of the visually selected active-BCG candidates described in Sec.~\ref{sec:init_visual}. The candidates exhibit a wide range of central colors, from cyan (1--12, Phoenix cluster), pink (4--6), white (1--5, Abell 1835), white-blue (3--3), purple (3--1), and orange-red (5--8). Many of these regions show spatially extended structure (e.g., 1--12, 2--3, 5--2, 6--2, 3--1, 3--3, 3--5, 4--6, 4--8, 5--8). 

Not all visually identified color deviations are due to activity within the BCG. The sample purity is reduced by line-of-sight projections with other types of objects. In the next subsection, we quantify the deviations in SB and color for the visual sample. Using these constraints, we then check the visual sample again for outliers in Sec. \ref{sec:gaia}.

\subsection{Color profiles and blue-core selection} \label{sec:colorprofiles}

\begin{figure*}
    \centering
    \includegraphics[width=\linewidth]{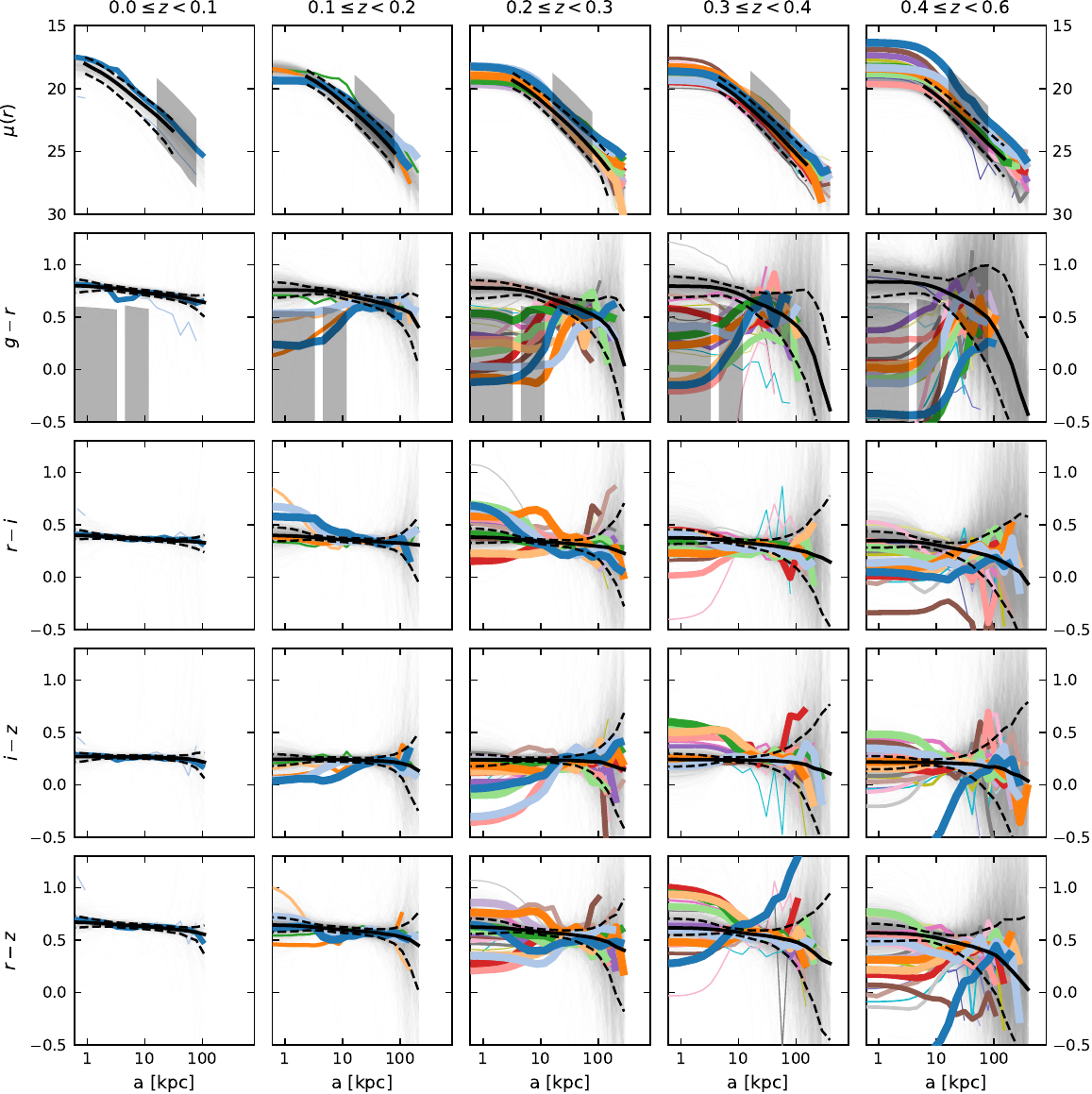}
    \caption{Radial SB and color profiles of BCGs in redshift bins. The top row shows the rest-frame $r$-band SB as a function of semi-major axis $a$, while the lower rows show rest-frame $g-r$, $r-i$, $i-z$, and $r-z$ color profiles. Light gray curves show individual parent-sample BCGs in the southern Legacy Surveys. Solid black curves show the median reference profiles, and dashed black curves mark the corresponding percentile envelopes: the 5th--95th percentiles for $\mu_r$ and the 16th--84th percentiles for the colors. Colored curves show the visually selected active-BCG candidates, with line width proportional to host-cluster X-ray luminosity. The thick blue curve in the highest-redshift bin is the Phoenix cluster BCG. 
    These candidates are distinguished primarily by a strong central $g-r$ blue excess within $a\lesssim20$\,kpc, whereas their $r$-band SB and redder color profiles more closely overlap with the parent population.
    Gray-shaded regions indicate the adopted color and SB selection windows.}
    \label{fig:visualcolor}
\end{figure*}

The radial SB and color profiles for most BCGs in our parent sample are shown in Fig. \ref{fig:visualcolor}. The columns are ordered by increasing redshift, with the top row showing the rest-frame $r$-band SB profiles and the subsequent rows showing rest-frame color profiles of progressively redder band combinations.

Individual light-gray profiles represent all parent BCGs with extended X-ray detections located in the southern Legacy Surveys.
The number of objects per redshift bin is 415, 1357, 1989, 2003, and 3586.
Continuous black lines show the median reference profiles, while dashed black lines show the corresponding percentile envelopes: 5th–95th percentiles for the SB profiles stacked at fixed SB levels, and 16th–84th percentiles for the color profiles computed in radial bins.

The median reference profiles exhibit negative color gradients, i.e., progressively bluer colors at larger radii, consistent with decreasing stellar metallicities in the outskirts (e.g., \citealt{Montes2014,Montes2018,Gu2020,Kluge2025}; Golden-Marx et al., submitted).
%,GoldenMarx2026}. % cite Jesse once his Paper is out
The scatter increases toward higher redshift, primarily because of larger photometric uncertainties and, possibly, residual uncertainties associated with the K-correction that is based on old stellar population templates, particularly for BCGs containing a young central stellar component. At large radii $a\sim100$\,kpc and high redshifts, the median and percentile envelopes show a pronounced downturn, driven by increasing uncertainties in the bluer bands.

Colored profiles show the visual sample. The line width is proportional to the host cluster X-ray luminosity. The number of objects per redshift bin is 2, 5, 19, 19, and 24. The thick blue line in the highest redshift bin is the Phoenix cluster's BCG.
The SB profiles are often, but not always, brighter in the center and the outskirts. Hence, the SB by itself is not a good discriminator between active and passive BCGs. The best discriminator appears to be the $g-r$ rest-frame color inside $a\lesssim20$\,kpc, where almost all BCGs in the visual sample are significantly bluer than the median BCG, typically by 0.5--1\,mag. Both the characteristic scale and color amplitude are consistent with the blue cores reported by \cite{Bildfell2008}. The redder color combinations show increased scatter but no, or much less, systematic differences between active and passive BCGs.

Using these findings, we tune constraints to automate the discovery of blue-core BCGs.
The shaded regions show the adopted color constraints. All data points at $a < 3.9$\,kpc must be bluer in $g-r$ than the reference median by at least 0.2\,mag. Between 4 and 15\,kpc, at least one data point must be bluer than the reference median by at least 0.15 mag. Between 15 and 80\,kpc, at least one data point must fall within the normal reference envelope, defined by the 16th–84th percentiles. In other words, the $g-r$ color profiles must be fully contained within the leftmost gray box and cross the middle and rightmost gray boxes.
Furthermore, over the radial range 15--80\,kpc, the $r$-band SB profile must remain entirely within the allowed SB interval: it must not be fainter than the median reference profile plus 1.4 times the median-to-95th-percentile offset, and not more than 3\,$r$\,mag brighter than the median reference profile. In other words, the ICL remains mostly unaffected by the central activity.

\begin{figure*}
    \centering
    \includegraphics[width=\linewidth]{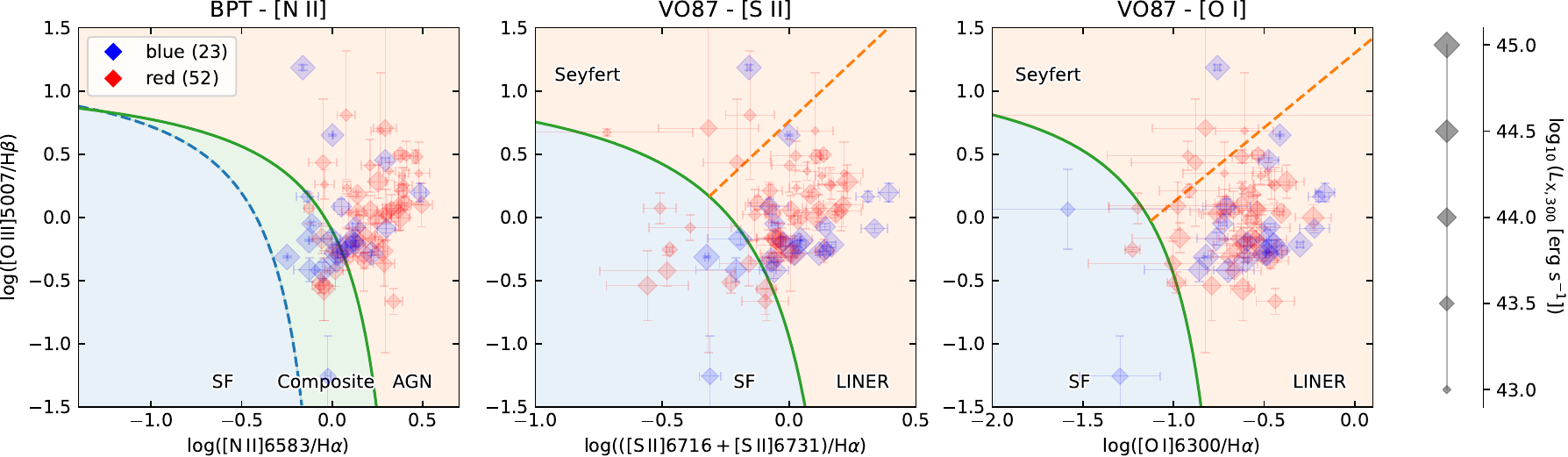}
    \caption{Optical emission-line diagnostic diagrams for BCGs with significant H$\alpha$ emission. From left to right, we show the classical [N\,{\sc ii}] BPT diagram and the [S\,{\sc ii}] and [O\,{\sc i}] VO87 diagrams. Blue and red symbols denote whether the BCG color profiles are classified as either blue or red, according to the cuts defined in Sec. \ref{sec:colorprofiles}. Error bars show the line-ratio uncertainties, and marker size encodes the host-cluster X-ray luminosity. The demarcation curves separate star-forming, composite, AGN, Seyfert, and LINER-like excitation regimes. Only narrow-line components are used. The numbers in the legend indicate the number of BCGs in each subsample.}
    \label{fig:bpt}
\end{figure*}

\subsection{Contamination and Gaia foreground-star rejection} \label{sec:gaia}

Not all visually identified color deviations are due to activity within the BCG. The candidates rejected by the quantitative color and SB constraints are marked by red frames in Fig. \ref{fig:visualsample}. Their causes are: 1--6, strong foreground Galactic extinction; 3--2, foreground star; 3--4, background galaxy; 3--8, source too compact to be a BCG; 3--9, normal BCG color profile; 4--2, foreground star; 4--3, gravitational lens; 4--4, foreground star; 4--12, source too compact to be a BCG; 5--1, incorrect BCG centering caused by a saturated foreground star; 5--3, source too compact to be a BCG; 5--6, tidal loop; 5--9, gravitational lens; 6--3, foreground star; 6--4, possible projected edge-on late-type cluster member; 6--8, possible projected late-type cluster member; and 6--11, source too compact to be a BCG.

Line-of-sight projections of foreground stars can mimic the appearance of active BCGs. The examples 4--1 and 3--10 illustrate this particularly well. They appear almost identical, while the former overlaps with a foreground star and the latter does not. Another similar example is 1--5 (Abell 1835), for which Gaia has falsely detected a proper motion with 3.2$\sigma$ significance but with a relatively high renormalized unit weight error (RUWE) of 1.37. There are further cases where Gaia stars were detected, but our color and SB rejection did not trigger: 2--8, 4--1, and 4--10.

To identify potential foreground stellar contamination, we cross-matched all blue-BCG positions against the Gaia Data Release 3 catalog \citep{Gaia2023} within $3\arcsec$ of the BCG center. We rejected candidates with a Gaia counterpart that showed both a significant proper motion, defined as an SNR greater than 3, and a robust astrometric point-source fit, quantified by $\mathrm{RUWE}<1.3$. The RUWE criterion prevents the rejection of genuine blue-BCG nuclei with non-stellar or variable central emission, for example from an AGN, which can produce spurious Gaia proper motions through centroid shifts or extended morphology.

The Gaia rejection is applied only to the visually selected candidates and blue-core BCGs, not to the red control samples. It removes three of the 69 visual candidates and 38 of the 420 color-selected BCGs. After additionally removing three secondary blue-core BCGs to avoid counting the same cluster twice, the final statistical sample includes 379 blue-core BCGs.

\subsection{Emission-line ionization diagnostics} \label{sec:BPT}

We use standard optical emission-line diagnostic diagrams to compare the excitation properties of blue BCGs and red BCGs with significant emission-line detections (see Sec. \ref{sec:fitspec}). Specifically, Fig. \ref{fig:bpt} shows the [N II]-based Baldwin–Phillips–Terlevich diagram \citep[BPT;][]{Baldwin1981}, together with the corresponding [S II] and [O I] Veilleux–Osterbrock diagrams \citep[VO87;][]{Veilleux1987}. The diagnostic regions are separated using the empirical star-forming/composite boundary of \citet{Kauffmann2003}, the theoretical maximum-starburst line of \citet{Kewley2001}, and the Seyfert/LINER (low-ionization nuclear emission-line region) divisions of \citet{Kewley2006}.
The plotted sample consists of 23 blue BCGs and 52 red BCGs with significant H$\alpha$ emission.

In the [N II] diagram, the blue BCGs are preferentially located toward, or within, the composite region between the star-forming and AGN demarcation curves. The red BCGs with significant H$\alpha$ emission lie mostly in the AGN region at higher $[\rm{N\ II}]/\rm{H}\alpha$. Thus, the two populations are mixed in line-ratio space, while the blue BCGs lie closer to the star-forming side of the diagnostic diagram.
In the [S II] and [O I] VO87 diagrams, the two subsamples largely overlap with no clear separation. Most objects from both subsamples lie outside the purely star-forming locus and are concentrated in the LINER region, with only a small number of sources extending toward the Seyfert regime. The X-ray luminosity is encoded by symbol size, showing a weak hint of higher $L_{\rm X,300}$ in the blue subsample.

\subsection{H$\alpha$ line emission and starburst tracks} \label{sec:ha_tracks}

\begin{figure*}
    \centering
    \includegraphics[width=\linewidth]{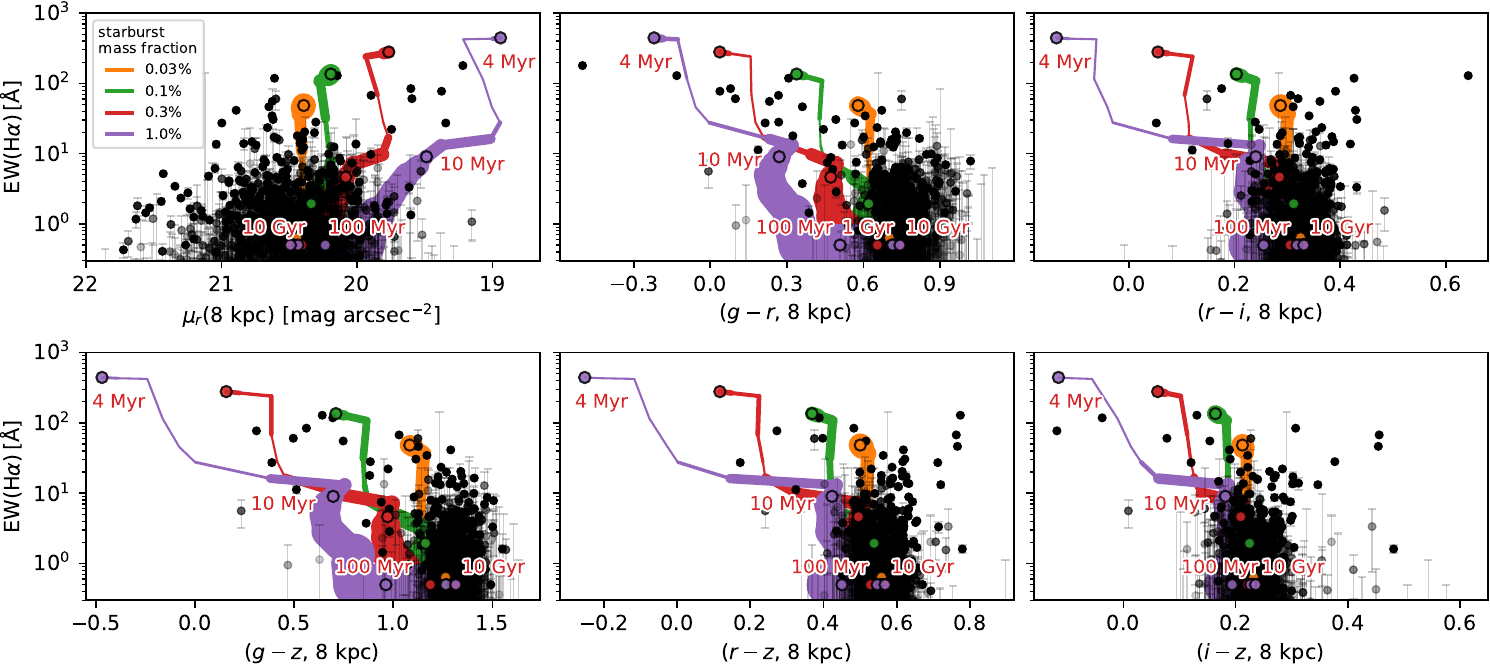}
    
    \caption{H$\alpha$ equivalent width compared to rest-frame $r$-band SB and rest-frame optical colors measured at $a=8\,{\rm kpc}$. Black circles show BCGs in the full spectroscopic sample, with opacity increasing with the SNR of the H$\alpha$ equivalent width; vertical error bars show the corresponding uncertainties. Colored tracks show FSPS/Prospector instantaneous-burst models in which a young stellar population is added to an old 10\,Gyr population, with burst mass fractions $f_{\rm burst}=0.03\%$, $0.1\%$, $0.3\%$, and $1.0\%$. Colored markers indicate selected burst ages, with age labels annotated for the $1\%$ track. The track width encodes the logarithmic residence-time density along the model trajectory. Model values below $\rm{EW(H\alpha)}=0.5$\,\AA\ are shown at this plotting floor. Stronger H$\alpha$ emission is associated with bluer $g-r$ (and $g-z$) colors, while the predicted accumulation of blue, H$\alpha$-weak post-starburst systems along the fading burst tracks is largely absent.}
    \label{fig:halpha_tracks}
\end{figure*}

Star formation is one process that can increase the brightness and shift the optical colors of BCGs toward bluer colors. As long as the starburst continues, it is accompanied by H$\alpha$ line emission. To investigate this relation, we utilize the full spectroscopic sample. We compare the line strength EW(H$\alpha$) to the rest-frame SB in the $r$ band and various rest-frame colors at $a=8$\,kpc in Fig. \ref{fig:halpha_tracks}.
We avoid the very center here because of the larger color uncertainties associated with the PSF convolution (see Sec. \ref{sec:meth:colorprofiles}). This choice has a negligible effect because the central color profiles are relatively flat.
Above $\rm{EW(H\alpha)}\gtrsim 10$\,\AA, a clear trend is seen in the $g-r$ color, where stronger H$\alpha$ emission correlates with bluer colors. A similar trend is seen in $g-z$ but not in the other colors. This confirms the finding in Sec. \ref{sec:colorprofiles} that active BCGs deviate mostly from passive BCGs in their $g-r$ color, that is, in the blue rest-frame continuum flux. The brightening in the $r$ band is only weak.

For comparison, we overplot the instantaneous-burst reference models introduced in Sec.~\ref{sec:starburst_models}. These tracks illustrate the expected joint evolution of the continuum colors and EW(H$\alpha$) in the idealized scenario of an isolated instantaneous starburst, and assuming that the H$\alpha$ emission is powered purely by star formation. Because the evolving young stellar population becomes progressively fainter relative to the old BCG, the model tracks evolve diagonally in color--EW(H$\alpha$) space rather than vertically at fixed color.

To visualize the relative time spent along the model trajectories, we vary the track width according to the logarithmic residence-time density, $\log_{10}(\Delta t/\Delta s)$, where $\Delta t$ is the elapsed time between adjacent native FSPS age points and $\Delta s$ is the distance traveled by the corresponding segment in the displayed plane. Thus, slowly evolving portions of a track are drawn more broadly than rapidly evolving portions. Once the model EW(H$\alpha$) falls below the plotting limit of 0.5\,\AA, it is shown at this floor and the linewidth is held fixed.
The even more strongly varying linear residence-time density, $\Delta t/\Delta s$, along the model track is proportional to the expected number density of fading post-starburst BCGs in the idealized scenario.
The models therefore predict a strong accumulation of objects in the slowly evolving, H$\alpha$-weak parts of the tracks at $g-r<0.6$ where the observed BCG number density increases at most mildly.

\subsection{X-ray properties of the host clusters}\label{sec:xray_properties}

\begin{figure*}
    \centering
    \includegraphics[width=0.49\linewidth]{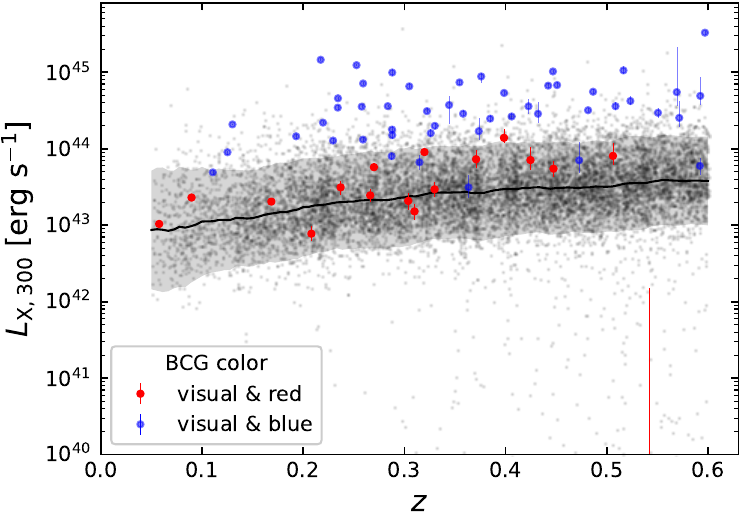}
    \includegraphics[width=0.49\linewidth]{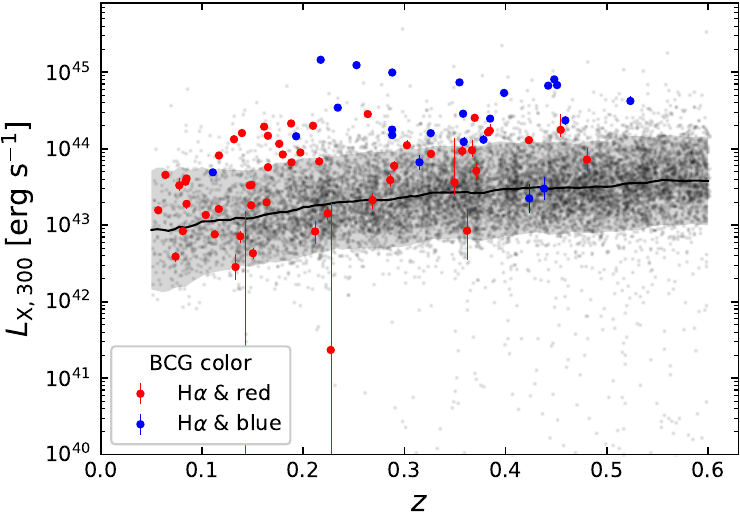}\\
    \includegraphics[width=0.49\linewidth]{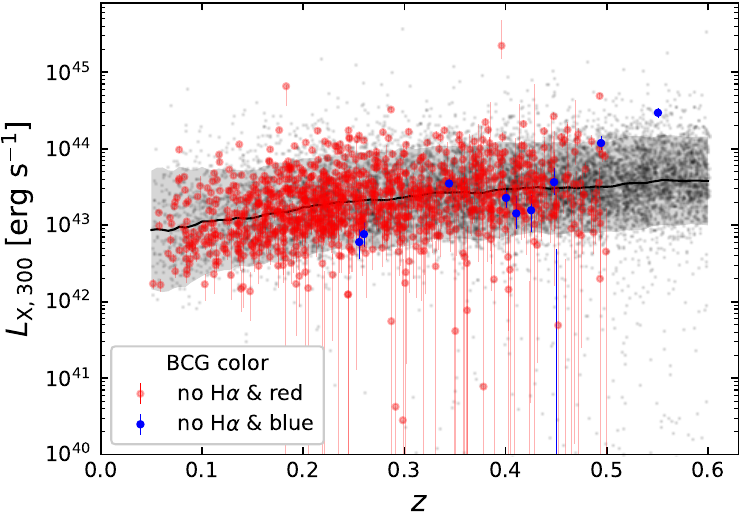}
    \includegraphics[width=0.49\linewidth]{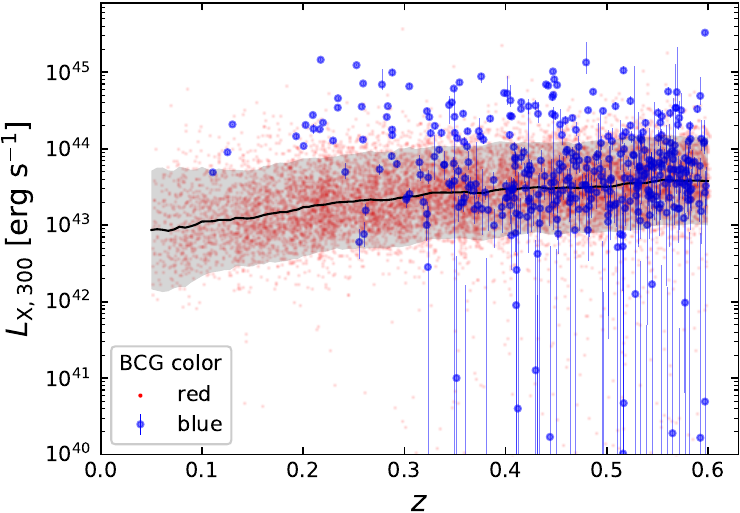}
    \caption{Host-cluster X-ray luminosity as a function of redshift for different BCG classifications. Gray points show the full statistical BCG sample. The black curve and gray shaded region show the running median and $\pm1.6\sigma_{\rm MAD}$ scatter. Highlighted points show, from top left to bottom right, visually selected candidates, H$\alpha$-emitting BCGs, BCGs without significant H$\alpha$ emission, and the full color-selected sample. Within each panel, blue and red symbols denote blue and red BCG classifications, respectively.
    The X-ray luminosity is measured in the 0.2--2.3\,keV band within 300\,kpc.
    The most X-ray-luminous cluster hosting a blue-core BCG in our sample is Phoenix at $z=0.597$, followed by RXCJ1504.1-0248 (RXJ1504) at $z=0.217$.}
    \label{fig:redshift_xray_host}
\end{figure*}

\begin{figure}
    \centering
    \includegraphics[width=\linewidth]{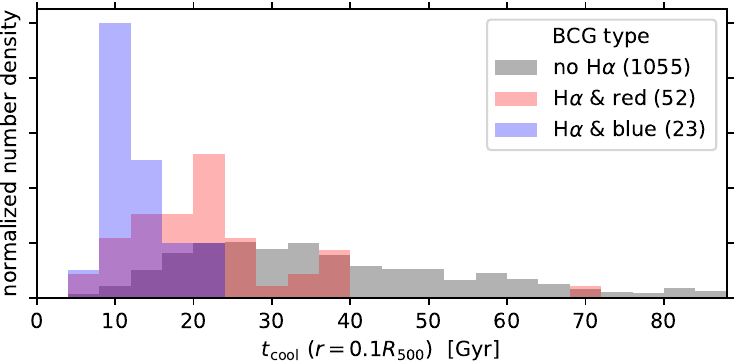}
    \includegraphics[width=\linewidth]{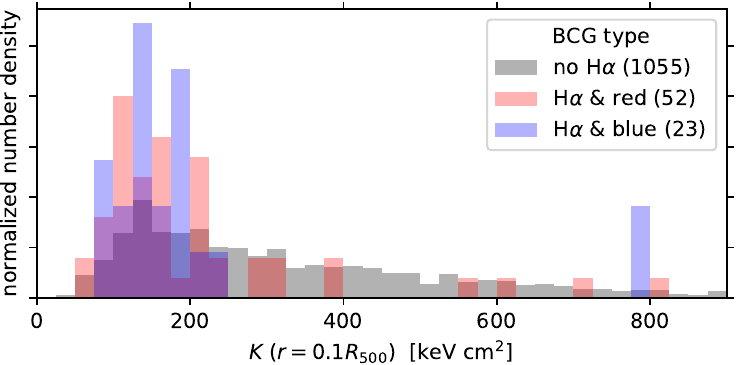}\\
    \caption{Thermodynamic properties of the spectroscopic BCG subsamples evaluated at $0.1R_{500}$. The normalized distributions of cooling time $t_{\rm cool}(0.1R_{500})$ and entropy $K(0.1R_{500})$ are shown for BCGs without significant H$\alpha$ emission, H$\alpha$-emitting red BCGs, and H$\alpha$-emitting blue BCGs. The numbers in the legend indicate the number of BCGs in each subsample.}
    \label{fig:histo_xray_host}
\end{figure}

\subsubsection{Luminosity} \label{sec:xray_luminosity}

We compare the optical properties of the BCGs to the X-ray properties of their host clusters in order to test whether blue colors and nebular emission preferentially occur in shorter-cooling, X-ray-luminous atmospheres. We use the 0.2--2.3\,keV X-ray luminosity measured within 300\,kpc, $L_{\rm X,300}$, together with entropy and cooling time evaluated at explicitly defined radii as described in Sec.~\ref{sec:xray_methods} and App. \ref{app:xray_radius}.
Figure~\ref{fig:redshift_xray_host} shows $L_{\rm X,300}$ as a function of redshift. The full statistical BCG sample is shown by the gray data points. The bulk of the population spans about two orders of magnitude in X-ray luminosity.
The black curve traces the running median of the parent population, while the shaded region indicates the corresponding $\pm1.6\sigma_{\rm MAD}$ scatter, where $\sigma_{\rm MAD}=1.4826\,{\rm MAD}$ and MAD is the median absolute deviation in $\log L_{\rm X,300}$, computed in redshift windows of width $\Delta z=0.05$. The factor 1.6 is chosen empirically such that the upper envelope approximately separates the visually selected blue and red BCG populations.

Visually selected blue BCGs are preferentially found at the high-luminosity end of the parent distribution, typically at $L_{\rm X,300}\gtrsim10^{44}\,{\rm erg\,s^{-1}}$, consistent with \cite{Green2016}, and usually near or above the upper scatter envelope. In contrast, objects classified as red BCGs within the visual sample occupy lower X-ray luminosities, mostly at $L_{\rm X,300}\lesssim10^{44}\,{\rm erg\,s^{-1}}$, supporting their interpretation as contaminants drawn from the passive parent population.

BCGs with central H$\alpha$ emission span a broad range of X-ray luminosities. Red H$\alpha$ emitters are mildly biased toward higher luminosities, whereas blue H$\alpha$ emitters show a much stronger preference for the high-luminosity end.

A complementary view is provided by the sample without significant H$\alpha$ detections. Most red BCGs in this sample follow the parent cluster population in the $L_{\rm X}$--$z$ plane with only one strong outlier, J114930.4+243928, above $L_{\rm X,300}\gtrsim10^{45}\,{\rm erg\,s^{-1}}$, whose X-ray luminosity is dominated by a blazar in a satellite galaxy. Blue BCGs without significant H$\alpha$ emission occur in only 10 systems, making them potential post-starburst candidates.

Purely color-selected blue BCGs show two distributions. One component resembles the visually selected blue-BCG sample at high luminosities. The other lies mostly within the scatter of the parent sample. These two components are approximately separated by the upper $1.6\sigma_{\rm MAD}$ envelope of the parent population. The relative contribution of the lower-luminosity component increases strongly above $z\approx0.3$. We do not expect a comparable increase in the intrinsic post-starburst fraction over this redshift range, whereas the photometric uncertainties increase substantially. We therefore interpret this second component primarily as contamination from intrinsically red BCGs that scatter into our blue color-selection constraints (see Fig.~\ref{fig:visualcolor}).
Restricting the color-selected sample to systems above the upper \(1.6\sigma_{\rm MAD}\) envelope leaves 98 blue BCGs, corresponding to \(98/11\,352 = 0.86\%\) of the statistical BCG sample.

\begin{table*}
\caption{Logistic-regression robustness tests for the richness dependence of the blue-core BCG fraction.}
\label{tab:logistic_richness}
\centering
\begin{tabular}{lllccc}
\hline\hline
Radius & Model & Sample & $N$ & $N_{\rm blue}$ & ${\rm OR}_{\lambda\times2}$ \\
\hline
$0.1R_{500}$ & linear & $t_{\rm cool}<20$ Gyr & 2208 & 122 & $2.45^{+0.39}_{-0.36}$ \\
$0.1R_{500}$ & quadratic & $t_{\rm cool}<20$ Gyr & 2208 & 122 & $2.52^{+0.43}_{-0.36}$ \\
$0.1R_{500}$ & linear & full valid sample & 11323 & 375 & $1.38^{+0.14}_{-0.12}$ \\
100 kpc & linear & $t_{\rm cool}<20$ Gyr & 1816 & 108 & $1.88^{+0.32}_{-0.28}$ \\
100 kpc & quadratic & $t_{\rm cool}<20$ Gyr & 1816 & 108 & $1.74^{+0.28}_{-0.24}$ \\
100 kpc & linear & full valid sample & 11323 & 375 & $1.46^{+0.15}_{-0.13}$ \\
\hline
\hline
\end{tabular}
\tablefoot{The radius specifies where $t_{\rm cool}$ is evaluated. ``Linear'' and ``quadratic'' refer to the adopted dependence on $\log t_{\rm cool}$ in the logistic-regression model; the quadratic model  includes an additional $(\log t_{\rm cool})^2$ term. The $t_{\rm cool}<20$ Gyr rows are restricted to the empirical shorter-cooling population, while the full-sample rows include all systems with a valid cooling-time measurement. $N$ and $N_{\rm blue}$ denote the total number of BCGs and the number classified as blue-core BCGs, respectively. ${\rm OR}_{\lambda\times2}$ is the odds ratio for a factor-of-two increase in richness after controlling for cooling time and redshift. Uncertainties are the 16th--84th percentile ranges from bootstrap resampling.}
\end{table*}

\subsubsection{Thermodynamic state at $0.1R_{500}$} \label{sec:thermodynamic_state}

At the fiducial radius of $0.1R_{500}$, motivated by the radial systematics analysis in App.~\ref{app:xray_radius}, the distributions of cooling time and entropy for the H$\alpha$-classified spectroscopic BCG sample are shown in Fig.~\ref{fig:histo_xray_host}.
The three spectroscopic populations show a clear ordering in cooling time, even though this radius does not directly probe the innermost cluster core, where the thermodynamic dependence of BCG activity is expected to be strongest. BCGs without significant H$\alpha$ emission span the broadest range of cooling times. Red H$\alpha$-emitting BCGs preferentially occupy shorter-cooling atmospheres, while blue H$\alpha$ emitters are most strongly concentrated toward the shortest cooling times.
In entropy, both H$\alpha$-emitting populations are more concentrated toward lower values than the population without significant H$\alpha$ emission.
A two-sample Kolmogorov--Smirnov test comparing the red and blue H$\alpha$ emitters confirms that their cooling-time distributions differ significantly ($D=0.43$, $p=0.003$), whereas the smaller separation of the entropy distributions ($D=0.21$) is not statistically significant ($p=0.41$).

\subsection{Blue-core incidence and dependence on redshift and richness}
\label{sec:incidence_richness}

\begin{figure*}
    \centering
    \includegraphics[width=0.49\linewidth]{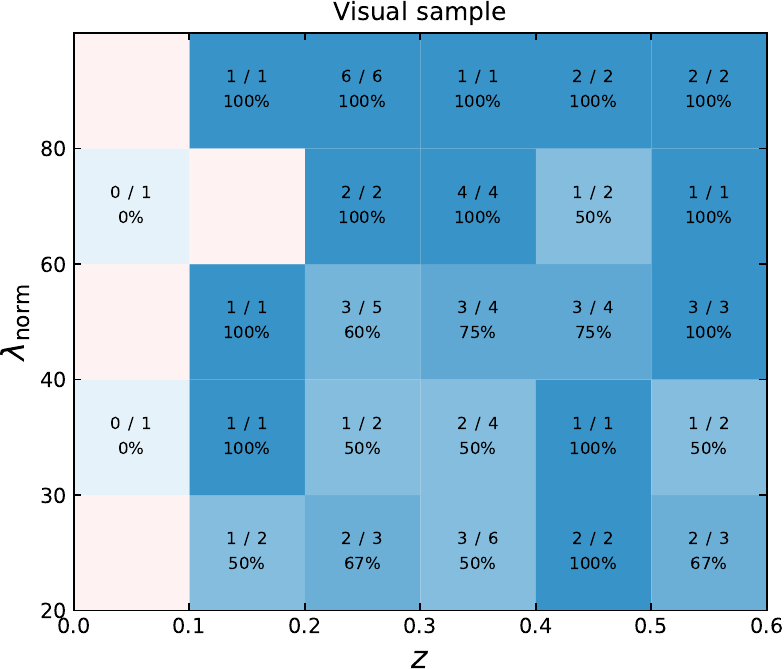}
    \includegraphics[width=0.49\linewidth]{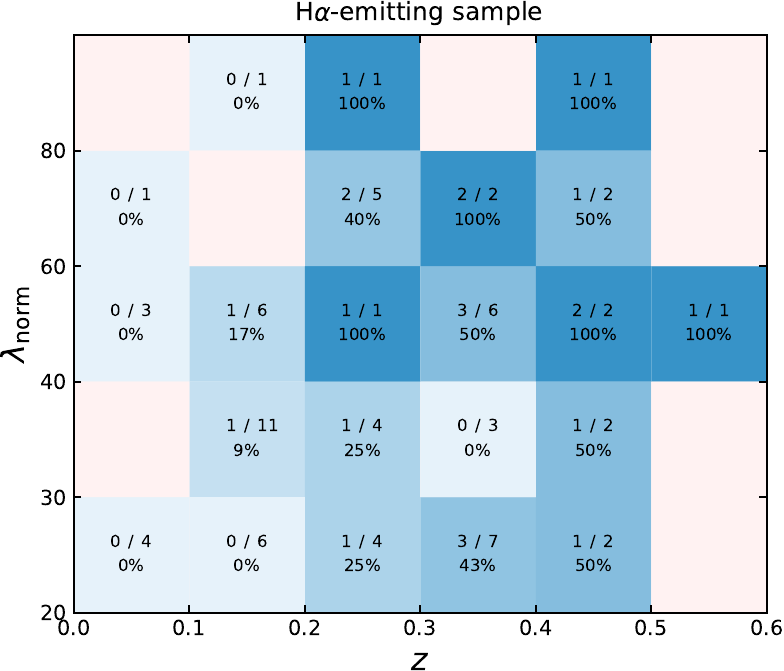}\\~\\
    \includegraphics[width=0.49\linewidth]{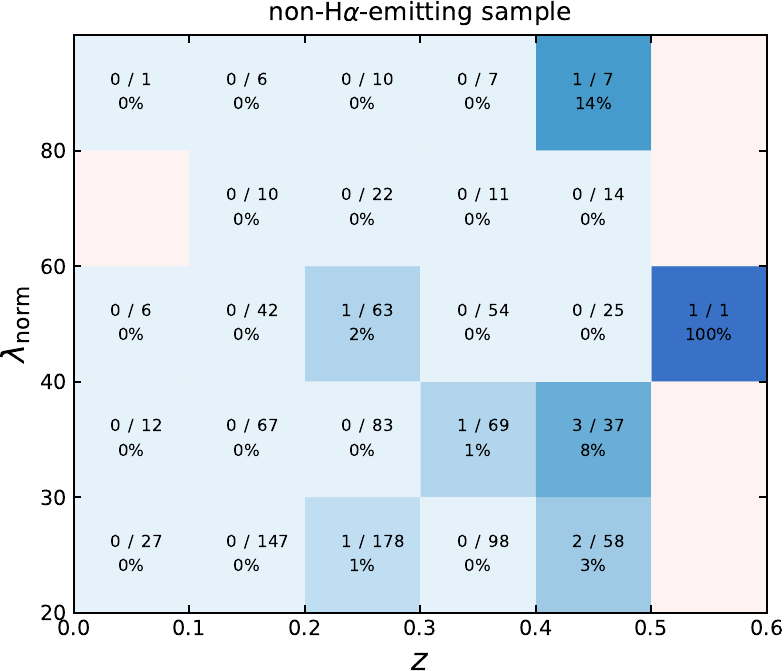}
    \includegraphics[width=0.49\linewidth]{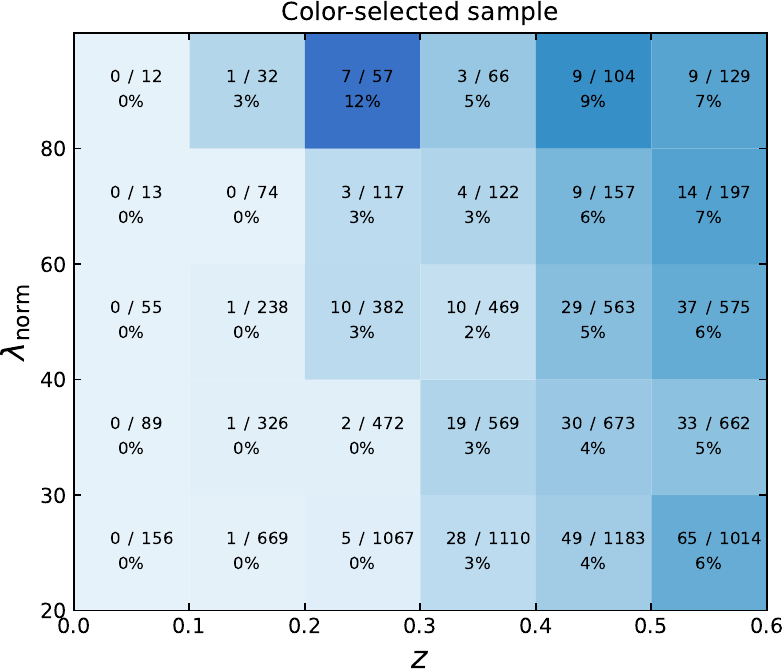}
    \caption{Incidence of blue BCGs in redshift--richness space. Each cell shows the number of blue BCGs and the total number of BCGs in that bin, along with the corresponding percentage; the cell color encodes the same fraction. The panels show, from top left to bottom right, the visually selected sample, the H$\alpha$-detected spectroscopic sample, the spectroscopic sample without significant H$\alpha$ emission, and the full color-selected statistical sample.}
    \label{fig:bluebcg_fraction}
\end{figure*}

Figure~\ref{fig:bluebcg_fraction} shows the fraction of blue BCGs as a function of redshift and cluster richness. We first consider the full color-selected statistical sample in the bottom-right panel, defined in Sec.~\ref{sec:statistical_sample}. Among the 11\,352 unique BCGs in this sample, 379 satisfy the blue-core criteria, corresponding to a total observed fraction of 3.3\%.

The color-selected blue-BCG fraction increases toward higher redshift. We do not interpret this trend as necessarily physical because the sensitivity of the color selection varies and the purity decreases with redshift. Although the selection is based mainly on K-corrected rest-frame $g-r$ color, the underlying photometry is measured in the observed-frame $g$ and $r$ bands, which sample different rest-frame wavelengths as a function of redshift. At low redshift, the observed $g$ band is less sensitive to the UV-bright young component, making weak blue cores more difficult to detect. At higher redshift, the BCGs become fainter in all observed bands, and the photometric uncertainties increase, particularly in the observed-frame $g$ band. The resulting scatter can move otherwise normal red BCGs into the fixed color-selection intervals at $a<15$\,kpc, as well as into the scatter-based color-selection region used for the ICL (see Fig.~\ref{fig:visualcolor}).

The dependence on richness in Fig.~\ref{fig:bluebcg_fraction}, bottom-right panel, is more robust because it can be assessed within individual redshift bins, where the observed-frame bandpass and typical photometric uncertainties are more similar. At fixed redshift, richer systems generally show higher blue-BCG fractions than poorer systems. This trend is most visible at intermediate and high redshifts. At $0.2<z<0.3$, the color-selected blue-BCG fraction rises from 0.5\% in the lowest-richness bin to 12\% in the highest-richness bin. Similarly, at $0.4<z<0.5$, the fraction increases from 4\% at low richness to about 9\% at high richness.

We also compare the full color-selected sample with the spectroscopic subsamples with and without significant H$\alpha$ emission. The H$\alpha$-emitting subsample in the top-right panel of Fig.~\ref{fig:bluebcg_fraction} shows substantially larger blue-BCG fractions, while the sample without significant H$\alpha$ emission in the bottom-left panel has much lower blue-BCG fractions. The visually selected sample in the top-left panel is intentionally biased toward conspicuous systems (see Sec. \ref{sec:init_visual}).

\begin{figure*}
    \centering
    \includegraphics[width=1\linewidth]{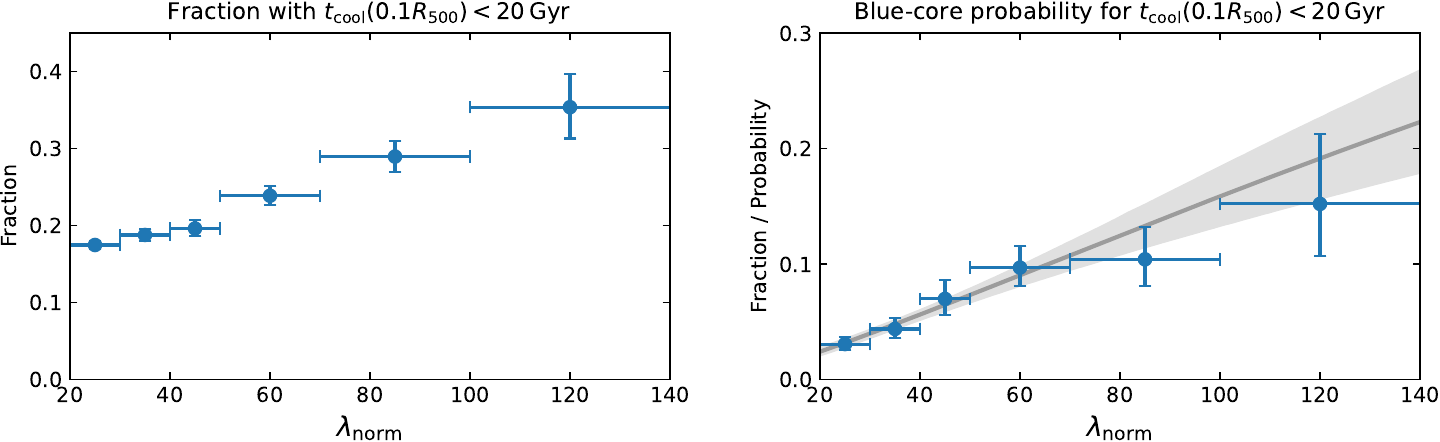}
    \caption{Dependence of ICM cooling and blue-core BCG incidence on cluster richness. Left: fraction of clusters with $t_{\rm cool}(0.1R_{500})<20$\,Gyr. 
    Right: observed blue-core fractions in richness bins and the standardized prediction of the logistic-regression model. For each richness, the model predictions are averaged over the observed population, thereby marginalizing over the observed $t_{\rm cool}$ and redshift distributions. The resulting modeled blue-core fraction is therefore directly comparable to the observed fractions, but is not a fit to the binned data. The shaded region shows the 16th--84th percentile bootstrap range.
    Horizontal error bars indicate the richness-bin widths, and vertical error bars show the 68\% Wilson binomial confidence intervals.}
    \label{fig:cool_fraction}
\end{figure*}

Figure~\ref{fig:cool_fraction} separates two related richness trends (left panel). The fraction of clusters satisfying $t_{\rm cool}(0.1R_{500})<20\,$Gyr increases moderately with richness. This trend is much steeper when cooling time is evaluated at a fixed physical radius due to the aperture effect, as discussed in App. \ref{app:xray_radius}. More importantly, among systems that already satisfy the shorter-cooling criterion, the blue-core fraction itself rises strongly with richness (right panel).
However, a thresholded fraction alone cannot establish that richness matters independently of the cooling-time and redshift distributions within the selected population. We therefore fit Eq.~\ref{eq:logistic_blue} to the individual clusters of the statistical BCG sample. The results are summarized in Tab.~\ref{tab:logistic_richness}. 
For the $t_{\rm cool}(0.1R_{500})<20\,$Gyr subset, 2208 systems enter the regression, including 122 blue-core BCGs.
The fitted richness coefficient is $b_\lambda=2.98^{+0.49}_{-0.52}$. Thus, after controlling statistically for the continuous measured cooling time and redshift, doubling richness increases the odds of hosting a blue-core BCG by a factor
\begin{equation}
{\rm OR}_{\lambda\times2}=2.45^{+0.39}_{-0.36}.
\end{equation}

To visualize the richness dependence implied by this multivariable fit, the right panel of Fig.~\ref{fig:cool_fraction} also shows the corresponding predicted blue-core probability. At each richness, we evaluate the model for every system using its observed $t_{\rm cool}$ and redshift and average the resulting predicted probabilities. This allows a direct comparison with the observed binned fractions. The predicted trend closely follows the increase seen in the observed binned fractions and is consistent with them within their statistical uncertainties.

The result is not tied to the richness-dependent definition of $0.1R_{500}$. Repeating the same analysis at a fixed radius of 100\,kpc gives 1816 shorter-cooling systems, including 108 blue-core BCGs, and yields $b_\lambda=2.10^{+0.53}_{-0.54}$, corresponding to ${\rm OR}_{\lambda\times2}=1.88^{+0.32}_{-0.28}$. Allowing a quadratic dependence on $\log t_{\rm cool}$ leaves the richness effect essentially unchanged: the odds ratios are $2.52^{+0.43}_{-0.36}$ at $0.1R_{500}$ and $1.74^{+0.28}_{-0.24}$ at 100\,kpc. 
We additionally repeat the linear regression without imposing the $t_{\rm cool}<20$\,Gyr selection. The full valid sample contains 11\,323 systems, including 375 blue-core BCGs, after removing 29 systems without valid cooling-time measurements. The richness dependence remains positive, with ${\rm OR}_{\lambda\times2}=1.38^{+0.14}_{-0.12}$ at $0.1R_{500}$ and $1.46^{+0.15}_{-0.13}$ at 100\,kpc. Thus, although the inferred richness dependence is stronger within the shorter-cooling population, it persists when longer-cooling systems are included. The positive richness trend is therefore not generated by the empirical 20\,Gyr selection. Taken together, the positive richness dependence survives changes in thermodynamic radius, a more flexible cooling-time dependence, and removal of the shorter-cooling selection.

A possible observational concern is that the detectability of a central blue excess might itself depend on richness if the central SB of the underlying passive stellar population changed strongly with cluster mass. We therefore examined the rest-frame $r$-band SB at $a=8\,{\rm kpc}$ as a function of richness. Within individual redshift intervals, the median SB brightens by only approximately $0.1$--$0.3\,r$ mag arcsec$^{-2}$ across the full richness range. The factor-of-several increase in the blue-core fraction is therefore unlikely to be caused solely by a richness-dependent sensitivity to central blue excesses.

\section{Discussion}\label{sec:discussion}

BCG activity is strongly associated with low-entropy, short-cooling cluster atmospheres \citep{Cavagnolo2008,Rafferty2008,Voit2008,Donahue2010,Pulido2018}, but favorable cooling conditions do not uniquely determine the state of the BCG. Short cooling times and close proximity to the X-ray peak have therefore been described as necessary but not sufficient conditions for recent star formation \citep{Loubser2016}. Our results extend this picture by statistically separating the roles of the thermodynamic state, cluster scale, and the different optical manifestations of BCG activity.

\subsection{Young stars and the origin of the optical signatures}
\label{sec:young_stars_interpretation}

The close relation between EW(H$\alpha$) and the central blue excess indicates that the blue continuum contains a young stellar component (Fig.~\ref{fig:halpha_tracks}). The instantaneous-burst tracks provide a reference for its approximate magnitude. Young components of order $0.1$--$1\%$ of the pre-existing stellar mass are sufficient to reproduce the prominent blue excesses, consistent with previous stellar-population studies that infer young stellar mass fractions from below one percent to a few percent in BCGs with recent star formation \citep[e.g.][]{Pipino2009,Liu2012,Loubser2016}.

The burst models should not, however, be interpreted as a unique decomposition of the observed emission. Most H$\alpha$-emitting BCGs do not occupy the pure star-forming locus in the BPT and VO87 diagrams (Fig.~\ref{fig:bpt}), and substantial LINER-like excitation is present in both the blue and red populations. Such low-ionization spectra are a well-established characteristic of line-emitting BCGs \citep[e.g.][]{Crawford1999,vonderLinden2007,Edwards2007,Loubser2013,Hamer2016}. The excitation mechanisms are diverse, with contributions from young stars, shocks, particle heating, and AGN activity \citep[e.g.][]{Hatch2007,McDonald2012,Hamer2016}. The shift of the blue H$\alpha$ emitters toward the composite region of the [N~{\sc ii}] BPT diagram is consistent with previous findings that blue or strongly star-forming BCGs show a larger stellar contribution to their nebular spectra \citep{Crawford1999,Hatch2007,Liu2012}.

Importantly, the non-stellar excitation mechanisms required by the emission-line ratios can contribute substantially to the nebular emission without producing a corresponding blue stellar continuum. Shocks, particle heating, and related mechanisms are invoked primarily to explain the line emission in BCG nebulae \citep[e.g.][]{Hatch2007,Hamer2016}, whereas the excess blue continuum observed in active BCGs has long been associated with young stellar populations and is independently supported by stellar absorption features and UV emission \citep{Cardiel1998,Crawford1999,Pipino2009,Loubser2016}. In Fig.~\ref{fig:halpha_tracks}, such additional line excitation would therefore primarily increase EW(H$\alpha$) without producing the accompanying shift toward bluer continuum colors. The blue excess consequently provides more direct evidence for a young stellar component, whereas the total H$\alpha$ emission cannot be attributed uniquely to star formation.

\subsection{Blue stellar components preferentially occur in the shortest-cooling atmospheres}
\label{sec:cooling_interpretation}

The association between BCG activity and rapidly cooling cluster atmospheres is well established. Optical line emission, blue or UV excesses, and recent star formation occur preferentially in systems with low central entropy and short cooling time, with characteristic scales of $K_0\sim30$\,keV\,cm$^2$ and $t_{\rm cool,0}\lesssim1$\,Gyr \citep{Cavagnolo2008,Rafferty2008,Voit2008,Donahue2010}. Similarly, \citet{Edwards2007} found the emission-line fraction to rise strongly in cooling-flow clusters and for BCGs located close to the X-ray center. These results established that a rapidly cooling, well-centered atmosphere provides favorable conditions for BCG activity.
More recently, \citet{Calzadilla2024} found that a low central entropy threshold associated with optical line emission persists out to $z\sim1$, with no significant evidence for redshift evolution.

The response of the BCG below these characteristic cooling scales is less uniform. Already \citet{Crawford1999} found that the most H$\alpha$-luminous BCGs generally show an excess blue continuum, whereas weaker line emitters can have continua similar to those of BCGs without line emission. \citet{Wang2010} likewise recovered the cooling-time threshold for H$\alpha$ emission and blue cores and found the two observables to be correlated, but reported substantial scatter and only a weak dependence of either quantity on the actual cooling time below the threshold.
However, \citet{Donahue2010} did find tentative trends of increasing H$\alpha$ equivalent width toward denser, shorter-cooling atmospheres in their cool-core cluster subsample. More direct evidence comes from infrared observations: \citet{ODea2008} reported increasing infrared luminosity toward shorter cooling times, and \citet{Rawle2012} found an anti-correlation between far-infrared-derived star-formation rate and X-ray cooling time among cool-core clusters.

Our population-level thermodynamic ordering is consistent with this picture. 
At $0.1R_{500}$, BCGs without significant H$\alpha$ emission span the broadest range of cooling times, red H$\alpha$ emitters preferentially occupy shorter-cooling atmospheres, and blue H$\alpha$ emitters are concentrated toward the shortest cooling times (Fig. \ref{fig:histo_xray_host}). A similar ordering is seen in $L_{\rm X,300}$, with red H$\alpha$ emitters mildly biased toward higher X-ray luminosities and blue H$\alpha$ emitters showing the strongest preference for X-ray-luminous systems (Fig. \ref{fig:redshift_xray_host}, top right panel).
These trends show that the probability of developing an optically conspicuous young stellar component continues to rise toward the most rapidly cooling systems within the short-cooling population.

\subsection{Cluster scale carries information beyond cooling time}
\label{sec:richness_interpretation}

At first sight, the positive richness dependence found in Sec.~\ref{sec:incidence_richness} (Figs.~\ref{fig:bluebcg_fraction} and \ref{fig:cool_fraction}) appears to differ from studies in which globally blue or star-forming BCGs occur preferentially in lower-mass systems. For example, \citet{Cerulo2019}, using integrated $g-r$ colors, found that the blue tail of the low-redshift BCG population is dominated by lower-mass clusters and groups, and that the star-forming fraction decreases with both cluster mass and BCG stellar mass.
A subsequent SED-based analysis of the same parent cluster catalog found only a weak dependence of the star-forming fraction on halo mass \citep{OrellanaGonzalez2022}.
More recently, \citet{Liu2025} reported similar trends using integrated $g-z$ colors, with star-forming BCGs also showing larger offsets from the optical cluster center. 
These populations, however, are not necessarily equivalent to the blue-core BCGs investigated here.
Selections based on integrated colors or general star-formation indicators can include late-type galaxies whose star formation need not be connected to ICM cooling. Such systems should be particularly common in lower-mass haloes, where the quenched fraction is lower than in massive cluster environments \citep{Wetzel2012,Woo2013}. 
Indeed, \cite{Pipino2011} noted that the enhanced blue-BCG fraction they found at low richness likely contains a contribution from spiral galaxies.
In contrast, our selection requires a centrally concentrated blue excess superimposed on an otherwise normal BCG profile together with a small separation from the X-ray center.
A more direct comparison is provided by \citet{Liu2012}, who found the incidence of emission-line and star-forming BCGs to increase from around the percent level at low richness up to approximately 5--10\% in the richest systems. This behavior is similar to the trend found here. However, their sample is broader than our centrally blue population. Our visual inspection classifies 16 of their 120 BCGs as late-type galaxies. 
Our result therefore confirms a positive richness trend for a population selected specifically for centrally concentrated blue-core activity in X-ray-selected clusters.

The logistic-regression analysis in Sec. \ref{sec:incidence_richness} sharpens this interpretation. Within the shorter-cooling population, richness remains a strong predictor of blue-core activity after cooling time and redshift are included as continuous covariates. Doubling richness increases the blue-core odds by $2.45^{+0.39}_{-0.36}$ at $0.1R_{500}$.
This shows that the measured cooling time alone does not capture all of the cluster-scale dependence of blue-core activity.

A natural physical interpretation is that cooling time and cluster scale describe different aspects of gas condensation. The cooling time is primarily an intensive quantity: it measures how rapidly gas at a given location can radiate its thermal energy, but not how much gas is available. By contrast, gas mass is an extensive quantity, such that a more massive atmosphere can provide a larger reservoir from which gas may cool or condense. This interpretation is qualitatively consistent with the increase of the upper envelope of the classical mass-deposition rate with cluster mass reported by \citet{Chen2007}, together with the relation between BCG star formation rate and absolute ICM cooling rate in strong cool cores \citep{McDonald2018,Calzadilla2022}.
The richness dependence may therefore reflect the larger gas reservoirs associated with larger-scale cluster atmospheres, and hence a larger absolute amount of gas available for cooling.

\subsection{The blue-core phase in the feedback-regulated cooling cycle}

The presented results suggest a picture in which the appearance of a blue stellar component is controlled by several aspects of the cooling atmosphere. Short cooling times provide the conditions under which condensation and BCG activity become increasingly likely, while the richness dependence indicates that the overall scale of the atmosphere carries additional information, plausibly through the amount of gas available to cool. At the same time, neither quantity uniquely determines the currently observed state of the BCG.
This diversity is seen even among apparently similar cool cores: \citet{McDonald2019a} found star-formation rates spanning a factor of $\sim$500 among four highly relaxed clusters despite broadly similar cooling times, cooling rates, and central entropies.
The observed populations may therefore be understood as different states of a feedback-regulated system in which the central thermodynamic state, the available gas reservoir, and the temporal phase of cooling and feedback jointly determine whether cooling leaves the BCG largely passive, produces line-emitting multiphase gas, or proceeds far enough to actively form stars.

The relative occurrence of H$\alpha$ emission and blue stellar light provides some information on this temporal dimension. Among the 1055 BCGs classified as having no significant H$\alpha$ emission, only ten are blue (Fig.~\ref{fig:redshift_xray_host}); four show tentative weak H$\alpha$ below our conservative detection threshold, four likely show no H$\alpha$, and two are too noisy to exclude weak emission. In the instantaneous-burst reference models, H$\alpha$ powered by young stars fades within $\sim$10\,Myr, whereas the blue continuum remains enhanced for $\sim$100\,Myr to $\sim$1\,Gyr (see Fig. \ref{fig:halpha_tracks}). A population dominated by isolated short bursts followed by long passive fading phases would therefore be expected to contain substantially more blue but H$\alpha$-weak systems than observed. Avoiding such a dominant post-starburst population would require star formation to remain active for much longer than the $\gtrsim100$\,Myr blue-continuum visibility time.
However, this conclusion is not unique because the line-ratio diagnostics (Fig. \ref{fig:bpt}) show that H$\alpha$ is not powered exclusively by young stars. Star formation could instead decline while other excitation mechanisms continue to power H$\alpha$. The blue continuum would then fade, and the BCG would return to the red H$\alpha$-emitting population without passing through a long-lived blue, H$\alpha$-weak phase.
Future high-SNR spectroscopy could distinguish these scenarios in two complementary ways: by searching for intermediate-age stellar populations in red H$\alpha$ emitters whose line emission may persist after star formation declines, and by testing a homogeneous non-H$\alpha$ sample for a post-starburst population after the line emission has faded.

The precipitation-regulated feedback framework \citep{Voit2015} provides a natural interpretation of these time-dependent changes in BCG activity.
Numerical models predict repeated cycles of cooling, cold-gas condensation, accretion, and AGN feedback \citep{Prasad2015}. Simulations that also include star formation connect these cycles to variations in the stellar component \citep{Li2015,Nobels2022}.
Importantly, the different components of this cycle do not evolve synchronously.
In the higher-mass cool-core systems simulated by \citet{Nobels2022}, decreasing cooling times precede enhanced star formation by $\sim$3$\times10^2$\,Myr. Star formation, in turn, precedes enhanced black-hole accretion, which subsequently increases the core's entropy and cooling time through AGN feedback. \cite{Li2015} showed that AGN feedback can suppress fresh ICM cooling while an existing cold-gas reservoir continues to sustain star formation for up to $\sim$1--2\,Gyr.
\cite{Prasad2015} likewise find hysteresis in the cold-gas/AGN cycle, although their simulations do not include star formation.
Consequently, clusters with similar observed cooling times and entropies can occupy different stages of the feedback cycle and exhibit very different levels of BCG activity. Such phase offsets provide a natural source of scatter in BCG activity at fixed thermodynamic state.

In this framework, the red H$\alpha$-emitting and blue H$\alpha$-emitting populations should not necessarily be interpreted as successive stages of a single deterministic sequence. A red H$\alpha$ emitter may develop a blue core during an episode of enhanced star formation and later return to redder colors, whereas other systems may never form enough young stars.
The blue-core population identified here can therefore be viewed as a rare and optically conspicuous phase of a broader cooling--feedback cycle: favorable thermodynamic conditions increase the likelihood of reaching this state, cluster scale provides an additional dependence, and the time-dependent response of condensation, star formation, and feedback determines whether a young stellar component is visible at the epoch of observation.

\section{Conclusions} \label{sec:conclusions}

We have presented a search for BCGs with centrally enhanced blue optical colors in the eRASS:5 cluster samples. Our selection identifies systems in which the BCG is closely aligned with the X-ray peak and shows a central color excess relative to the normal red BCG population. Combining radial SB and color profiles, optical spectroscopy, and eROSITA-derived X-ray properties, we investigated how the thermodynamic state and scale of the cluster atmosphere are related to the appearance of an optically conspicuous young stellar component. Our main conclusions are as follows.

\begin{enumerate}

\item Blue-core BCGs are rare: we identify 379 among 11\,352 systems, corresponding to an observed incidence of 3.3\%. The apparent increase with redshift is likely dominated by changing sensitivity and increasing photometric contamination. Restricting to the 98 blue-core BCGs above the upper X-ray luminosity envelope defines a higher-purity subset comprising 0.86\% of the full sample.

\item The defining photometric signature is a central $g-r$ color at least 0.2\,mag bluer than the reference population, typically extending to $a\lesssim20$\,kpc before returning to the normal BCG envelope. Its correlation with EW(H$\alpha$) indicates a young stellar component, with burst mass fractions of order 0.1--1\% of the pre-existing stellar mass in our reference models.

\item H$\alpha$ emission is closely linked to the blue continuum but is not a pure star-formation tracer. Blue H$\alpha$ emitters lie preferentially toward the composite region of the [N~{\sc ii}] BPT diagram, while both blue and red populations show substantial LINER-like excitation. The blue continuum therefore provides more direct evidence for young stars than H$\alpha$ emission alone.

\item The spectroscopic populations show a clear ordering in cooling time at $0.1R_{500}$: non-H$\alpha$ BCGs span the broadest range, red H$\alpha$ emitters preferentially occupy shorter-cooling atmospheres, and blue H$\alpha$ emitters are most strongly concentrated toward the shortest cooling times. Both H$\alpha$-emitting populations are also concentrated toward lower entropies than the non-H$\alpha$ population. The red and blue H$\alpha$ cooling-time distributions differ significantly, whereas their entropy distributions do not.

\item Blue-core BCGs preferentially occur in X-ray-luminous clusters. Visually confirmed blue systems and blue H$\alpha$ emitters are typically found at $L_{\rm X,300}\gtrsim10^{44}\,{\rm erg\,s^{-1}}$. Red H$\alpha$ emitters are also biased toward higher X-ray luminosities.

\item Cluster scale carries information about blue-core activity beyond the measured cooling time. For systems with $t_{\rm cool}(0.1R_{500})<20$\,Gyr, doubling the richness increases the blue-core odds by $2.45^{+0.39}_{-0.36}$ after controlling for cooling time and redshift, and the positive trend persists under all robustness tests. This may reflect the larger gas reservoirs and hence larger absolute cooling or condensation rates associated with larger-scale, generally more massive cluster atmospheres.

\item The blue stellar component appears to represent a restricted phase within a broader line-emitting multiphase state. The scarcity of blue BCGs without significant H$\alpha$ emission is inconsistent with a population dominated by isolated short bursts followed by long H$\alpha$-quiet fading phases. It can instead be explained either by sustained or closely recurrent star formation over timescales much longer than $100$\,Myr, or by H$\alpha$ emission persisting through additional excitation mechanisms as the young stellar component fades.

\end{enumerate}

Taken together, these results identify centrally blue BCGs as a rare and optically conspicuous manifestation of feedback-regulated ICM cooling. Favorable thermodynamic conditions increase the likelihood of reaching this state, while the additional richness dependence indicates that the overall scale of the cluster atmosphere also matters. Neither quantity uniquely determines the instantaneous appearance of the BCG. Instead, the observed diversity is consistent with a cooling--feedback cycle in which the central thermodynamic state, the available gas reservoir, and the time-dependent response of condensation, star formation, and feedback jointly determine whether cooling leaves the BCG largely passive, produces line-emitting multiphase gas, or proceeds far enough to actively form stars.

% ----------------------------------------------------------------------

\begin{acknowledgements}
%We thank the anonymous referee for constructive comments.
%\\
% ERC
E. Bulbul, S. Zelmer, E. Artis, and X. Zhang acknowledge financial support from the European Research Council (ERC) Consolidator Grant under the European Union’s Horizon 2020 research and innovation program (grant agreement CoG DarkQuest No 101002585). 
\\
% eROSITA
This work is based on data from eROSITA, the soft X-ray instrument aboard SRG, a joint Russian-German science mission supported by the Russian Space Agency (Roskosmos), in the interests of the Russian Academy of Sciences represented by its Space Research Institute (IKI), and the Deutsches Zentrum f{\"{u}}r Luft und Raumfahrt (DLR). The SRG spacecraft was built by Lavochkin Association (NPOL) and its subcontractors and is operated by NPOL with support from the Max Planck Institute for Extraterrestrial Physics (MPE).
\\
The development and construction of the eROSITA X-ray instrument were led by MPE, with contributions from the Dr. Karl Remeis Observatory Bamberg \& ECAP (FAU Erlangen-Nuernberg), the University of Hamburg Observatory, the Leibniz Institute for Astrophysics Potsdam (AIP), and the Institute for Astronomy and Astrophysics of the University of T{\"{u}}bingen, with the support of DLR and the Max Planck Society. The Argelander Institute for Astronomy of the University of Bonn and the Ludwig Maximilians Universit{\"{a}}t Munich also participated in the science preparation for eROSITA.
\\
The eROSITA data shown here were processed using the eSASS/NRTA software system developed by the German eROSITA consortium.
\\
% Legacy Surveys
The Legacy Surveys consist of three individual and complementary projects: the Dark Energy Camera Legacy Survey (DECaLS; Proposal ID \#2014B-0404; PIs: David Schlegel and Arjun Dey), the Beijing-Arizona Sky Survey (BASS; NOAO Prop. ID \#2015A-0801; PIs: Zhou Xu and Xiaohui Fan), and the Mayall z-band Legacy Survey (MzLS; Prop. ID \#2016A-0453; PI: Arjun Dey). DECaLS, BASS and MzLS together include data obtained, respectively, at the Blanco telescope, Cerro Tololo Inter-American Observatory, NSF’s NOIRLab; the Bok telescope, Steward Observatory, University of Arizona; and the Mayall telescope, Kitt Peak National Observatory, NOIRLab. Pipeline processing and analyses of the data were supported by NOIRLab and the Lawrence Berkeley National Laboratory (LBNL). The Legacy Surveys project is honored to be permitted to conduct astronomical research on Iolkam Du’ag (Kitt Peak), a mountain with particular significance to the Tohono O’odham Nation.
\\
NOIRLab is operated by the Association of Universities for Research in Astronomy (AURA) under a cooperative agreement with the National Science Foundation. LBNL is managed by the Regents of the University of California under contract to the U.S. Department of Energy.
\\
This project used data obtained with the Dark Energy Camera (DECam), which was constructed by the Dark Energy Survey (DES) collaboration. Funding for the DES Projects has been provided by the U.S. Department of Energy, the U.S. National Science Foundation, the Ministry of Science and Education of Spain, the Science and Technology Facilities Council of the United Kingdom, the Higher Education Funding Council for England, the National Center for Supercomputing Applications at the University of Illinois at Urbana-Champaign, the Kavli Institute of Cosmological Physics at the University of Chicago, Center for Cosmology and Astro-Particle Physics at the Ohio State University, the Mitchell Institute for Fundamental Physics and Astronomy at Texas A\&M University, Financiadora de Estudos e Projetos, Fundacao Carlos Chagas Filho de Amparo, Financiadora de Estudos e Projetos, Fundacao Carlos Chagas Filho de Amparo a Pesquisa do Estado do Rio de Janeiro, Conselho Nacional de Desenvolvimento Cientifico e Tecnologico and the Ministerio da Ciencia, Tecnologia e Inovacao, the Deutsche Forschungsgemeinschaft and the Collaborating Institutions in the Dark Energy Survey. The Collaborating Institutions are Argonne National Laboratory, the University of California at Santa Cruz, the University of Cambridge, Centro de Investigaciones Energeticas, Medioambientales y Tecnologicas-Madrid, the University of Chicago, University College London, the DES-Brazil Consortium, the University of Edinburgh, the Eidgenossische Technische Hochschule (ETH) Zurich, Fermi National Accelerator Laboratory, the University of Illinois at Urbana-Champaign, the Institut de Ciencies de l’Espai (IEEC/CSIC), the Institut de Fisica d’Altes Energies, Lawrence Berkeley National Laboratory, the Ludwig Maximilians Universitat Munchen and the associated Excellence Cluster Universe, the University of Michigan, NSF’s NOIRLab, the University of Nottingham, the Ohio State University, the University of Pennsylvania, the University of Portsmouth, SLAC National Accelerator Laboratory, Stanford University, the University of Sussex, and Texas A\&M University.
\\
BASS is a key project of the Telescope Access Program (TAP), which has been funded by the National Astronomical Observatories of China, the Chinese Academy of Sciences (the Strategic Priority Research Program “The Emergence of Cosmological Structures” Grant \# XDB09000000), and the Special Fund for Astronomy from the Ministry of Finance. The BASS is also supported by the External Cooperation Program of Chinese Academy of Sciences (Grant \# 114A11KYSB20160057), and Chinese National Natural Science Foundation (Grant \# 12120101003, \# 11433005).
\\
The Legacy Survey team makes use of data products from the Near-Earth Object Wide-field Infrared Survey Explorer (NEOWISE), which is a project of the Jet Propulsion Laboratory/California Institute of Technology. NEOWISE is funded by the National Aeronautics and Space Administration.
\\
The Legacy Surveys imaging of the DESI footprint is supported by the Director, Office of Science, Office of High Energy Physics of the U.S. Department of Energy under Contract No. DE-AC02-05CH1123, by the National Energy Research Scientific Computing Center, a DOE Office of Science User Facility under the same contract; and by the U.S. National Science Foundation, Division of Astronomical Sciences under Contract No. AST-0950945 to NOAO.
\\
% DESI
This research used data obtained with the Dark Energy Spectroscopic Instrument (DESI). DESI construction and operations is managed by the Lawrence Berkeley National Laboratory. This material is based upon work supported by the U.S. Department of Energy, Office of Science, Office of High-Energy Physics, under Contract No. DE–AC02–05CH11231, and by the National Energy Research Scientific Computing Center, a DOE Office of Science User Facility under the same contract. Additional support for DESI was provided by the U.S. National Science Foundation (NSF), Division of Astronomical Sciences under Contract No. AST-0950945 to the NSF’s National Optical-Infrared Astronomy Research Laboratory; the Science and Technology Facilities Council of the United Kingdom; the Gordon and Betty Moore Foundation; the Heising-Simons Foundation; the French Alternative Energies and Atomic Energy Commission (CEA); the National Council of Humanities, Science and Technology of Mexico (CONAHCYT); the Ministry of Science and Innovation of Spain (MICINN), and by the DESI Member Institutions: www.desi.lbl.gov/collaborating-institutions. The DESI collaboration is honored to be permitted to conduct scientific research on I’oligam Du’ag (Kitt Peak), a mountain with particular significance to the Tohono O’odham Nation. Any opinions, findings, and conclusions or recommendations expressed in this material are those of the author(s) and do not necessarily reflect the views of the U.S. National Science Foundation, the U.S. Department of Energy, or any of the listed funding agencies.
% SDSS
Funding for the Sloan Digital Sky Survey V has been provided by the Alfred P. Sloan Foundation, the Heising-Simons Foundation, the National Science Foundation, and the Participating Institutions. SDSS acknowledges support and resources from the Center for High-Performance Computing at the University of Utah. SDSS telescopes are located at Apache Point Observatory, funded by the Astrophysical Research Consortium and operated by New Mexico State University, and at Las Campanas Observatory, operated by the Carnegie Institution for Science. The SDSS web site is \url{www.sdss.org}.
\\
SDSS is managed by the Astrophysical Research Consortium for the Participating Institutions of the SDSS Collaboration, including the Carnegie Institution for Science, Chilean National Time Allocation Committee (CNTAC) ratified researchers, Caltech, the Gotham Participation Group, Harvard University, Heidelberg University, The Flatiron Institute, The Johns Hopkins University, L'Ecole polytechnique f\'{e}d\'{e}rale de Lausanne (EPFL), Leibniz-Institut f\"{u}r Astrophysik Potsdam (AIP), Max-Planck-Institut f\"{u}r Astronomie (MPIA Heidelberg), Max-Planck-Institut f\"{u}r Extraterrestrische Physik (MPE), Nanjing University, National Astronomical Observatories of China (NAOC), New Mexico State University, The Ohio State University, Pennsylvania State University, Smithsonian Astrophysical Observatory, Space Telescope Science Institute (STScI), the Stellar Astrophysics Participation Group, Universidad Nacional Aut\'{o}noma de M\'{e}xico, University of Arizona, University of Colorado Boulder, University of Illinois at Urbana-Champaign, University of Toronto, University of Utah, University of Virginia, Yale University, and Yunnan University.  
\\
% HET
The Hobby-Eberly Telescope (HET) is a joint project of the University of Texas at Austin, the Pennsylvania State University, Ludwig-Maximilians-Universit\"at M\"unchen, and Georg-August Universit\"at G\"ottingen. The HET is named in honor of its principal benefactors, William P. Hobby and Robert E. Eberly.

\end{acknowledgements}

\section*{Disclosure of Delegation to Generative AI}

The authors declare the use of generative AI in the research and writing
process. According to the GAIDeT taxonomy \citep{Suchikova2026}, the following tasks were
delegated to GAI tools under full human supervision:

\begin{itemize}
    \item Literature search and systematization
    \item Code generation
    \item Code optimization
    \item Data analysis
    \item Visualization
    \item Text generation
    \item Proofreading and editing
    \item Summarizing text
    \item Formulation of conclusions
    \item Quality assessment
    \item Identification of limitations
    \item Recommendations
    \item Publication support
\end{itemize}

\noindent The GAI tool used was OpenAI ChatGPT (GPT-5.6 and GPT-6).

\noindent Responsibility for the final manuscript lies entirely with the authors.
GAI tools are not listed as authors and do not bear responsibility for
the final outcomes.

\noindent Declaration submitted by: Matthias Kluge

\bibliographystyle{aa}
\bibliography{ref_bcg}

\begin{appendix}

\section{Training and diagnostics of the BCG ranker} \label{app:bcg}

The BCG ranker introduced in Sec.~\ref{sec:bcg1} is used to identify BCGs in the forthcoming eRASS:5 cluster catalog. Here, we provide additional background on the BCG-identification criteria, the construction of the visual training sample, and diagnostics of the trained model.

Although the term brightest cluster galaxy (BCG) literally refers to the most luminous galaxy in a cluster, in practice BCG identification also relies on its location near the center of the cluster potential and its disproportionately extended morphology relative to the size--luminosity relation of normal early-type galaxies \citep[e.g.,][]{Kluge2020}. These criteria do not always identify the same galaxy, making the BCG assignment ambiguous in $\sim$20\% of clusters \citep{Kluge2020}. X-ray information significantly improves the centering criterion because the peak of the X-ray-emitting, virialized ICM traces the deepest point in the cluster potential well, particularly in cool-core clusters.

Guided by these criteria, we constructed the training sample from a reevaluation of the 4303 eRASS1 clusters visually inspected by \citet{Zenteno2025}, selected primarily with $\lambda_{\rm norm}>16$ and $M_{500}>10^{14}\,{\rm M_\odot}$. After matching to eRASS:5 and adding systems at higher redshift, we visually identified the BCG in 5283 clusters using primarily its location relative to the X-ray center and its extended morphology. The training set includes several dozen active BCGs identified during this inspection.

\begin{figure}
    \centering
    \includegraphics[width=\linewidth]{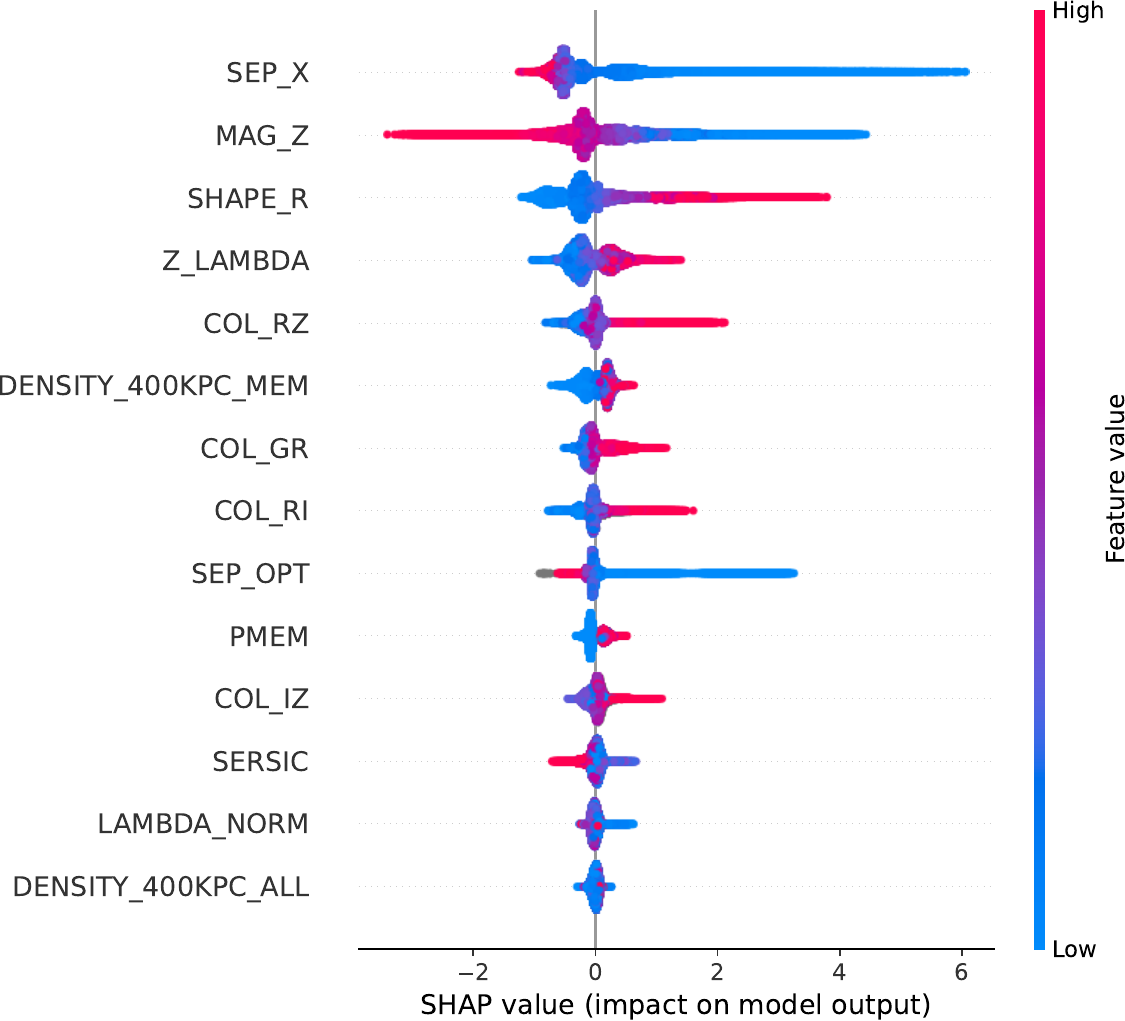}
    \caption{SHAP summary plot for the LightGBM BCG ranker. Each point represents one candidate galaxy, and the horizontal axis gives the SHAP contribution to the model score for identifying the galaxy as the BCG. Positive SHAP values increase the BCG ranking score, while negative values decrease it.}
    \label{fig:shap}
\end{figure}

Figure~\ref{fig:shap} shows the relative importance of the input features. Large absolute SHAP values indicate a strong impact on the BCG classification, with positive values increasing and negative values decreasing the model score. At the top of the figure are the features with the highest impact: angular separation between the X-ray detection and the galaxy ({\tt SEP\_X}), extinction-corrected galaxy magnitude in the $z$ band ({\tt MAG\_Z}), and angular half-light radius of the galaxy.
Low feature values (blue), corresponding to proximity for {\tt SEP\_X} and bright $z$-band magnitudes for {\tt MAG\_Z}, increase the likelihood of being the BCG, as does a high feature value of {\tt SHAPE\_R}, i.e., a large galaxy angular extent.
A red $r-z$ color ({\tt COL\_RZ}), which is least noisy at high redshifts compared to other colors, also increases the likelihood. However, the constraining power is lower than that of the top three features, thereby also allowing correct identification of active BCGs with deviating colors. A high density of eROMaPPer-selected members within a radius of 400\,kpc around the X-ray center {\tt DENSITY\_400KPC\_MEM} indicates proximity to the cluster center, where galaxy density is typically highest.
The remaining, less impactful features include the cluster photometric redshift ({\tt Z\_LAMBDA}) and richness ({\tt LAMBDA\_NORM}), the cluster membership probability of the galaxy under consideration ({\tt PMEM}), its separation from the optical cluster center ({\tt SEP\_OPT}), additional galaxy colors ({\tt COL\_GR}, {\tt COL\_RI}, {\tt COL\_IZ}), the S\'ersic index ({\tt SERSIC}), and the projected density of all galaxies within a radius of 400\,kpc around the X-ray center ({\tt DENSITY\_400KPC\_ALL}).

\section{Spectral fitting and emission-line uncertainties} \label{app:spectral_fitting}

In this section, we provide the technical details of the spectral fitting and emission-line uncertainty estimation summarized in Sec.~\ref{sec:fitspec}.
For each spectrum, the procedure is as follows:
\begin{enumerate}
\item the input spectrum is logarithmically rebinned;
\item the stellar templates are convolved to match the instrumental resolution of the observed spectrum (using \texttt{WDISP} with the correction described in \cite{KlugeBender2023} for SDSS, 1.7\,\AA~full width at half maximum for DESI, and the channel-dependent resolutions determined in \cite{KlugeBender2023} for HET/LRS2), divided by $1+z$,
\item the stellar continuum is fitted with \texttt{pPXF} \citep{Cappellari2004,Cappellari2017} as a linear combination of stellar templates without additive or multiplicative polynomials to avoid biasing emission-line fluxes;
\item narrow emission-line templates are fit simultaneously with the continuum using Gaussian line templates generated at the mean instrumental resolution of the spectrum, with velocity dispersions restricted to $\sigma < 800$\,km\,s$^{-1}$;
\item if needed, an additional broad Balmer-line component is included, with velocity dispersions restricted to $\sigma > 800$\,km\,s$^{-1}$;
\item the noise is estimated in ten wavelength segments from the residuals of an initial fit using the median absolute deviation, interpolated over wavelength, and scaled up by a factor of 1.5;
\item the final fit is repeated using this noise estimate, including the stellar continuum and gas components simultaneously.
\end{enumerate}

For each fitted emission line, the integrated flux $F_{\rm line}$ is computed from the fitted \texttt{pPXF} gas-template weight $w_{\rm line}$ and the wavelength-integrated template profile,
\begin{equation}
    F_{\rm line} = w_{\rm line} \sum_i G_i\,\Delta\lambda_i ,
\end{equation}
where $G_i$ is the Gaussian emission-line template value at logarithmic wavelength pixel $i$, and $\Delta\lambda_i$ is the corresponding linear wavelength width. Doublet lines are fitted with two Gaussians and a tied theoretical amplitude ratio of 3:1 for [O\,III] $\lambda5007/\lambda4959$, [O\,I] $\lambda6300/\lambda6364$, and [N\,II] $\lambda6583/\lambda6548$.

Equivalent widths are computed from the fitted line template relative to the best-fitting stellar continuum,
\begin{equation}
    {\rm EW}_{\rm line} = w_{\rm line} \sum_i \frac{G_i\,\Delta\lambda_i}{C_i},
\end{equation}
where $C_i$ is the continuum model at pixel $i$, obtained from the best-fitting model after subtracting the fitted gas components. The line template and continuum are evaluated in the same normalized flux-density units, so the absolute flux normalization cancels and the equivalent width is obtained in \AA. Since the calculation is performed on the rest-frame logarithmic wavelength grid, the reported equivalent widths are rest-frame equivalent widths.

Narrow and broad components are measured separately. For a given transition, the above expressions are applied independently to the narrow and, where present, broad gas-template columns, with their respective fitted weights $w_{\rm line}$ and template profiles $G_i$. The resulting narrow-component fluxes and equivalent widths are stored separately from the broad Balmer-component measurements.

The line-flux uncertainty is not returned directly by \texttt{pPXF}. We therefore estimate it from the weighted linear least-squares uncertainty of the fitted gas-template amplitude,
\begin{equation}
    \sigma_w \simeq \left[ \sum_i \left(\frac{G_i}{\sigma_i}\right)^2 \right]^{-1/2} \sqrt{\chi^2_{\rm red}},
\end{equation}
where $\sigma_i$ is the spectral noise used in the final fit and $\chi^2_{\rm red}$ is the reduced chi-square of the full \texttt{pPXF} model. The corresponding integrated-flux and equivalent-width uncertainties are
\begin{equation}
    \sigma_F = \sigma_w \sum_i G_i\,\Delta\lambda_i ,
\end{equation}
and
\begin{equation}
    \sigma_{\rm EW} = \sigma_w \sum_i \frac{G_i\,\Delta\lambda_i}{C_i}.
\end{equation}

For a fixed Gaussian line template and a locally valid noise vector, this expression gives the standard statistical uncertainty of the fitted line amplitude. The factor $\sqrt{\chi^2_{\rm red}}$ rescales the formal uncertainty to account for a residual mismatch between the adopted noise normalization and the scatter of the final-fit residuals. The calculation uses the spectral noise vector at the line position; it does not add an extra Poisson term from the emission-line photons themselves. Covariance with neighboring emission lines, stellar continuum templates, and sky residuals is also neglected.

\section{Radial systematics and fiducial thermodynamic radius}
\label{app:xray_radius}

\begin{figure*}
    \centering
    \includegraphics[width=\textwidth]{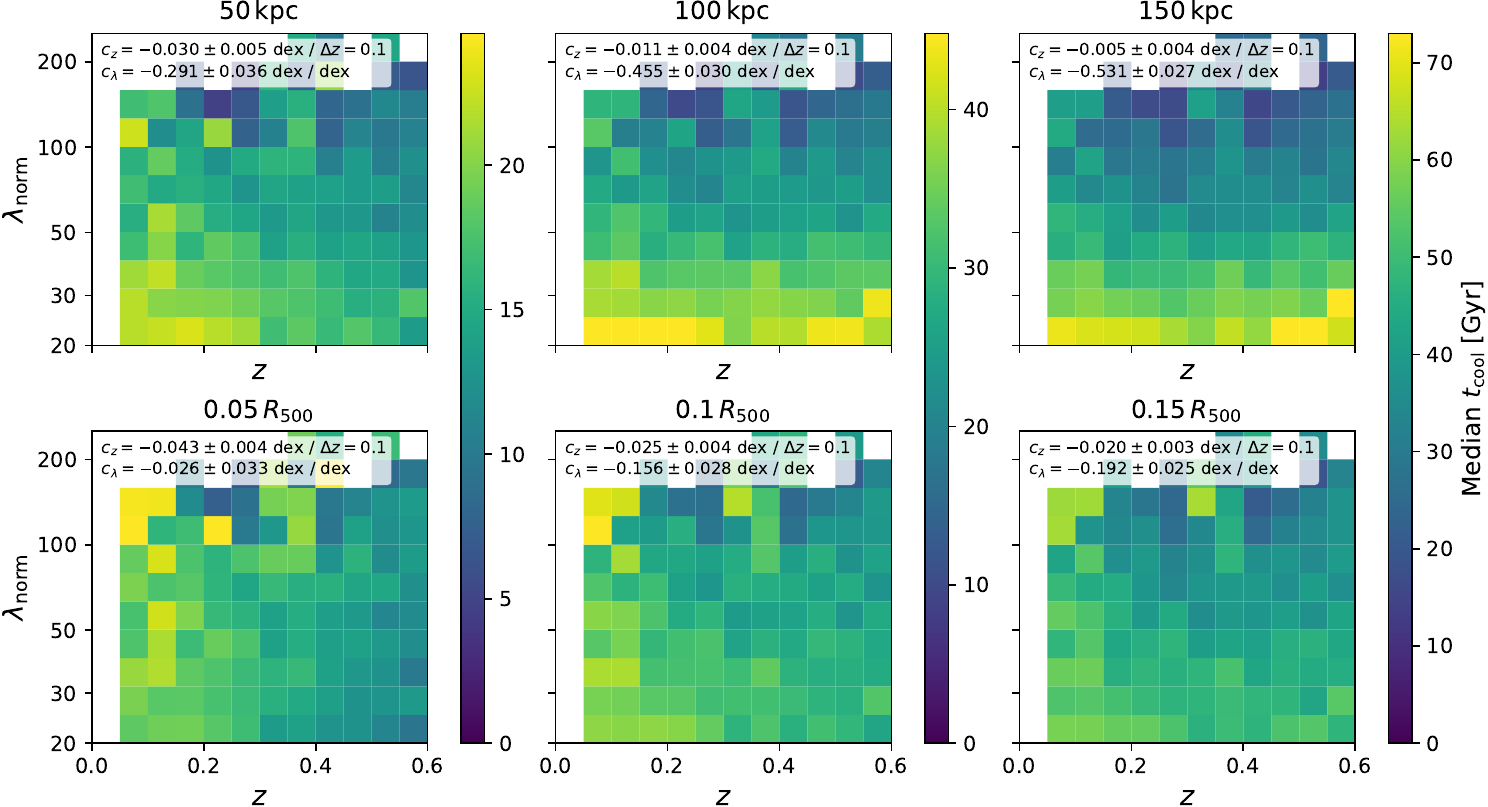}
    \caption{Median cooling times of the extended-source subset of the statistical BCG sample in the redshift--richness plane. The profiles are evaluated at 50, 100, and 150\,kpc (top row) and at $0.05$, $0.10$, and $0.15R_{500}$ (bottom row). The annotations show the fitted redshift coefficient $c_z$ per $\Delta z=0.1$ and richness coefficient $c_\lambda$ per dex from Eq.~\ref{eq:tcool_radial_trends}.}
    \label{fig:tcool_radial_systematics}
\end{figure*}

In this section, we motivate our choice of $0.1R_{500}$ as our fiducial radius.
The cooling-time trends examined in Sec. \ref{sec:xray_methods} depend substantially on the radius at which the ICM profile is evaluated. Figure~\ref{fig:tcool_radial_systematics} shows the median cooling times of the extended-source subset of the statistical BCG sample in redshift--richness bins. We evaluate the profiles at three fixed physical radii and three radii scaled by $R_{500}$.
It reveals two distinct trends: a horizontal trend with redshift and a vertical trend with richness. Their dependence on the adopted radius is complementary. The redshift trend becomes strongest at small radii, where limited spatial resolution and inward profile extrapolation are most important, whereas the richness trend becomes strongest at fixed physical radii, where the same aperture probes systematically different fractions of $R_{500}$ across cluster mass.

To quantify these trends, at each radius we fit
\begin{equation}
\log_{10}\!\left(\frac{t_{\rm cool}}{{\rm Gyr}}\right)
=
a_t
+ c_z \frac{z-0.3}{0.1}
+ c_\lambda \log_{10}\!\left(\frac{\lambda_{\rm norm}}{50}\right).
\label{eq:tcool_radial_trends}
\end{equation}
The fit is performed to the median values in the redshift--richness bins. Coefficient uncertainties are estimated as half the 16th--84th percentile interval from 1000 bootstrap resamplings of the clusters. The annotations in Fig.~\ref{fig:tcool_radial_systematics} give the resulting redshift and richness coefficients, $c_z$ and $c_\lambda$, respectively.

At fixed physical radii, the apparent redshift dependence becomes progressively weaker toward larger radius, from $c_z=-0.030\pm0.005$ dex per $\Delta z=0.1$ at 50\,kpc to $-0.005\pm0.004$ at 150\,kpc. Such a bias is expected due to the decreasing eROSITA spatial resolution at higher redshift and the inward extrapolation of the fitted thermodynamic profiles \cite[e.g.,][]{Main2017}. Although the eROSITA PSF is included in the forward modeling, this does not fully restore spatial information on scales unresolved by the data. Moreover, selection effects and genuine redshift evolution may also contribute to the trend.

The choice of radius also strongly affects the apparent richness dependence. At 100\,kpc, $c_\lambda=-0.455\pm0.030$ dex per dex, whereas at $0.1R_{500}$ it is only $-0.156\pm0.028$ dex per dex. This is expected because a fixed physical radius probes a smaller fraction of $R_{500}$ in richer systems and therefore samples systematically farther inward in scaled radius.

Scaled radii substantially reduce this mass-dependent aperture effect, although they retain some residual redshift dependence. We therefore adopt $0.1R_{500}$ as our fiducial thermodynamic radius. 
This choice is a compromise between reducing the richness-dependent aperture effect and avoiding the stronger sensitivity to resolution and extrapolation apparent at $0.05R_{500}$, while remaining close to the cluster core.
Although we focus on richness trends in this work and do not interpret the redshift trends physically, minimizing the latter remains important because they indicate that the $t_{\rm cool}$ measurements themselves become unreliable.
As a complementary fixed-radius robustness test at a comparable physical scale, we use 100\,kpc, which is close to the median $0.1R_{500}=86$\,kpc in our statistical BCG sample.

% \section{Catalog of active BCGs.} \label{sec:app_catalog}

\end{appendix}

\end{document}